\documentclass[aps,prl,superscriptaddress,reprint]{revtex4-1}
\usepackage{amsmath}
\usepackage{amssymb}
\usepackage{graphicx}
\usepackage[colorlinks,urlcolor=blue]{hyperref}
\usepackage{url}
\usepackage{color,xcolor}
\usepackage{ulem}
\usepackage{float}
\usepackage{epstopdf}
\begin{document}

\title{Magnetic states of quasi-one-dimensional iron chalcogenide Ba$_2$FeS$_3$}
\author{Yang Zhang}
\author{Ling-Fang Lin}
\affiliation{Department of Physics and Astronomy, University of Tennessee, Knoxville, Tennessee 37996, USA}
\author{Gonzalo Alvarez}
\affiliation{Computational Sciences \& Engineering Division and Center for Nanophase Materials Sciences, Oak Ridge National Laboratory, Oak Ridge, TN 37831, USA}
\author{Adriana Moreo}
\author{Elbio Dagotto}
\affiliation{Department of Physics and Astronomy, University of Tennessee, Knoxville, Tennessee 37996, USA}
\affiliation{Materials Science and Technology Division, Oak Ridge National Laboratory, Oak Ridge, Tennessee 37831, USA}

\begin{abstract}
Quasi-one-dimensional iron-based ladders and chains, with the 3$d$ iron electronic density $n = 6$, are attracting considerable attention.
Recently, a new iron chain system Ba$_2$FeS$_3$, also with $n = 6$, was prepared under high-pressure and high-temperature conditions. Here, the magnetic and electronic phase diagrams are theoretically studied for
this quasi-one-dimensional compound. Based on first-principles calculations, a strongly anisotropic
one-dimensional electronic band behavior near the Fermi level was observed. In addition, a three-orbital
electronic Hubbard model for this chain was constructed. Introducing the Hubbard and Hund
couplings and studying the model via the density matrix renormalization group (DMRG) method, we studied the ground-state phase diagram. A robust staggered $\uparrow$-$\downarrow$-$\uparrow$-$\downarrow$ AFM region
was unveiled in the chain direction, consistent with our density functional theory (DFT) calculations. Furthermore, at intermediate Hubbard $U$ coupling strengths, this system was found to display an orbital selective Mott phase (OSMP)
with one localized orbital and two itinerant metallic orbitals. At very large $U/W$ ($W$ = bandwidth), the system displays Mott insulator characteristics, with two orbitals half-filled and one doubly occupied.
Our results for high pressure Ba$_2$FeS$_3$ provide guidance to experimentalists and theorists working on this one-dimensional iron chalcogenide chain material.
\end{abstract}

\maketitle

\section{I. Introduction}
One-dimensional material systems continue attracting considerable attention due to their rich physical properties, where the charge, spin, orbital, and lattice degrees of freedom are intertwinned in a reduced dimensional phase space~\cite{Yunoki:prl,Dagotto:rmp94,Grioni:JPCM,Monceau:ap}. Remarkable physical phenomena have been reported in different one-dimensional systems, such as high critical temperature superconductivity in copper chains and ladders~\cite{cu-ladder1,cu-ladder2,cu-ladder3}, ferroelectricity~\cite{Park:prl,Lin:prm,Zhang:prb21}, spin block states~\cite{Zhang:prbblock,Herbrych:prbblock}, nodes in spin density ~\cite{Lin:prb21}, orbital ordering ~\cite{Pandey:prb21,Lin:prm21}, orbital-selective Mott phases~\cite{Patel:osmp,Herbrych:osmp}, and charge density wave or Peierls distortions~\cite{Zhang:prb21,Zhang:prbcdw,Gooth:nature,Zhang:arxiv}.

Because superconductivity at high pressure was reported a few years ago in the two-leg ladder compounds BaFe$_2$$X_3$ ($X$ = S, Se)~\cite{Takahashi:Nm,Ying:prb17} with electronic density $n = 6$, the iron ladders became interesting one-dimensional systems to research high-temperature iron-based superconductors~\cite{Yamauchi:prl15,Zhang:prb17,Zhang:prb18,Zheng:prb18,Zhang:prb19,Zhang:prb19,Materne:prb19,Pizarro:prm,Wu:prb19,Craco:prb20}. BaFe$_2$S$_3$ displays a stripe-type antiferromagnetic (AFM) order below $12$~K, involving AFM legs and ferromagnetic (FM) rungs (this state is called CX)~\cite{Takahashi:Nm,chi:prl}. In addition, BaFe$_2$Se$_3$, namely replacing S by Se, displays an exotic AFM state with $2\times2$ FM blocks coupled antiferromagnetically along the long ladder direction below $256$~K under ambient conditions~\cite{Caron:Prb,Caron:Prb12}.  By applying hydrostatic pressure, both systems display an insulator-metal transition~\cite{Zhang:prb17,Zhang:prb18,Materne:prb19,Ying:prb17}, followed by superconductivity at $P$ $\sim 11$ Gpa~\cite{Takahashi:Nm,Ying:prb17}. Furthermore, an OSMP state was found in BaFe$_2$Se$_3$ by neutron experiments at ambient pressure~\cite{mourigal:prl15}. This state was theoretically predicted before experimental confirmation by using the density matrix renormalization group (DMRG) method based on a multi-orbital Hubbard model~\cite{osmp1,osmp2}. These developments in the area of two-leg iron ladder systems naturally introduce a simple question: can iron chains, as opposed to ladders, also display similar physical properties?

Some iron chalcogenide chains $A$Fe$X_2$ ($A$ = K, Rb, Cs and Tl, $X$ = S or Se) have already been prepared experimentally~\cite{Seidov:prb01,Seidov:prb16}. Neutron diffraction experiments revealed that the magnetic coupling along the chain direction is AFM with dominant $\pi$ wavevector ($\uparrow$-$\downarrow$-$\uparrow$-$\downarrow$)~\cite{Bronger:jssc,Seidov:prb16}, similar to the CX-AFM state in BaFe$_2$S$_3$. But in $A$Fe$X_2$ compounds there are 5 electrons in the $3d$ iron orbitals, corresponding to valence Fe$^{\rm 3+}$. At this electronic density $n = 5$, the AFM phase with the $\uparrow$-$\downarrow$-$\uparrow$-$\downarrow$ configuration was observed in a large portion of the magnetic phase diagram when using the five-orbital Hubbard model for iron chains studied via the real-space Hartree-Fock (HF) approximation~\cite{Luo:prb14}. However, these HF calculations reported a much richer phase diagram for iron chains at electronic density $n = 6$. Considering that the atomic electronic density $n=6$ is the same as in iron planar and ladder
superconductors~\cite{Dagotto:Rmp}, then iron chains with electronic density $n = 6$ could be potential candidates to achieve a similar superconducting state.

Na$_2$FeSe$_2$ with $n=6$ was considered as a candidate~\cite{Stuble:jssc}. Recent DMRG calculations~\cite{Pandey:prb} found a stable region of staggered spin order in the phase diagram (with $\uparrow$-$\downarrow$-$\uparrow$-$\downarrow$  order) at low Hund coupling $J_H$/U, while block phases ($\uparrow$-$\uparrow$-$\downarrow$-$\downarrow$) dominate at larger $J_H$/U. Another interesting iron chain with $n=6$ $Ln$$_2$O$_2$FeSe$_2$ ($Ln$ = Ce, La) was prepared experimentally~\cite{McCabe:cc,McCabe:prb}. In addition, OSMP and Hund physics were discussed in this compound by using DMRG-based calculations based on the Hubbard model~\cite{Lin:osmp}.

\begin{figure}
\centering
\includegraphics[width=0.48\textwidth]{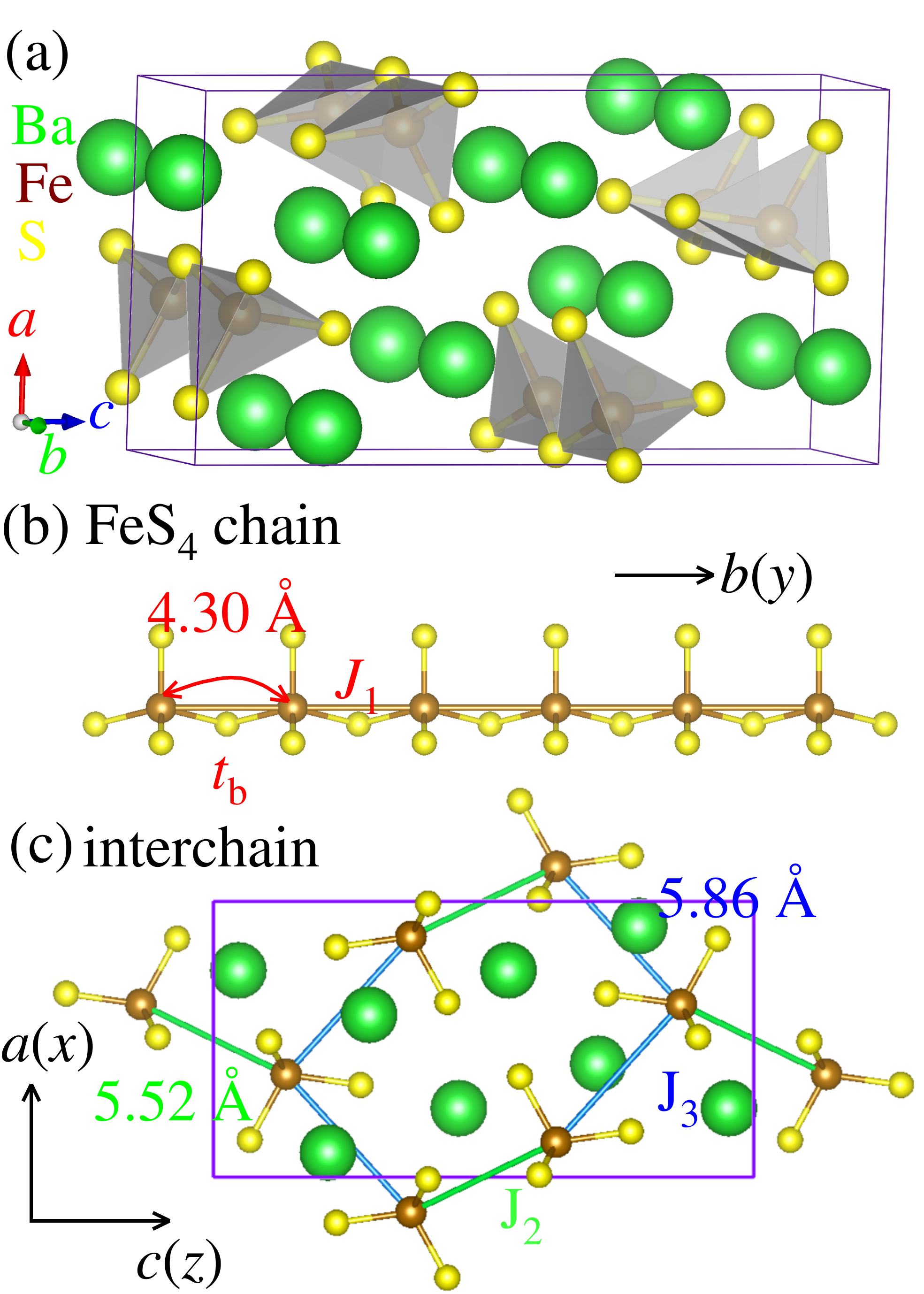}
\caption{Schematic crystal structure of the high-pressure Ba$_2$FeS$_3$ conventional cell (green = Ba; brown = Fe; yellow = S). (a) Sketch of the $ac$ plane along $b$ direction. (b) FeS$_4$ chain along the $b$-axis. (c) Inter chains magnetic exchange couplings on the $ac$ plane. }
\label{Fig1}
\end{figure}

Very recently, a new $n=6$ iron chalcogenide Ba$_2$FeS$_3$ (note this is a 213 formula, unlike the 123 of ladders) was synthesized under high-pressure (HP) and high-temperature conditions~\cite{Guan:jac}. A long-range AFM transition was reported at $\sim 56$~K, and the magnetic susceptibility curve exhibited a round hump behavior until $110$~K~\cite{Guan:jac}. As shown in Fig.~\ref{Fig1}(a), Ba$_2$FeS$_3$ (HP) has an orthorhombic structure with space group $Pnma$ (No. $62$). In the crystal structure of  Ba$_2$FeS$_3$ (HP), there are four FeS$_4$ chains along the $b$-axis, where nearest-neighbor irons are connected by sulfur atoms along the chain direction [see  Fig.~\ref{Fig1}(b)]. The nearest-neighbor (NN) Fe-Fe bond is $4.30$~\AA, corresponding to the lattice constant along the $b$-axis ~\cite{Guan:jac}. In addition, the NN and next-nearest-neighbor (NNN) interchain distances are $5.52$~\AA ~and $5.86$~\AA, respectively, as displayed in Fig.~\ref{Fig1}(c). Based on the crystal structure, the Ba$_2$FeS$_3$ (HP) phase displays quasi one-dimensional characteristics, suggesting
that the chain direction plays the dominant role in the physical properties.

To better understand the electronic and magnetic structures, here both first-principles DFT and DMRG methods were employed to investigate Ba$_2$FeS$_3$ at high pressure. First, the {\it ab initio} DFT calculations indicated a strongly anisotropic
electronic structure for Ba$_2$FeS$_3$ (HP), in agreement with its anticipated one-dimensional geometry. Furthermore, based on DFT calculations, we found the staggered spin order was the most likely magnetic ground state, with the coupling along the chain direction dominanted by the $\pi$ wavevector. Based on the Wannier functions obtained from first-principle calculations, we obtained the relevant hopping amplitudes and on-site energies for the iron atoms. Next, we constructed a multi-orbital Hubbard model for the iron chains. Based on the DMRG calculations, we calculated the ground-state phase diagram varying the on-site Hubbard repulsion $U$ and the on-site Hund coupling $J_H$. The staggered AFM with $\uparrow$-$\downarrow$-$\uparrow$-$\downarrow$ was found to be dominant in a robust portion of the phase diagram, in agreement with DFT calculations. In addition, OSMP physical properties were found in the regime of intermediate Hubbard coupling strengths. Eventually, at very large $U/W$, the OSMP is replaced by the Mott insulating (MI) phase.

\section{II. Method}

\subsection{A. DFT Method}
In the present study, the first-principles DFT calculations were performed with the projector augmented wave (PAW) method, as implemented in the Vienna {\it ab initio} simulation package (VASP) code~\cite{Kresse:Prb,Kresse:Prb96,Blochl:Prb}. Here, the electronic correlations were considered by using the generalized gradient approximation (GGA) with the Perdew-Burke-Ernzerhof  functional~\cite{Perdew:Prl}. The plane-wave cutoff was $500$ eV. The k-point mesh was $8\times16\times4$ for the non-magnetic calculations, which was accordingly adapted for the magnetic calculations. Note that we tested explicitly that this $k$-point mesh already leads to converged energies. Furthermore, the local spin density approach (LSDA) plus $U_{\rm eff}$ with the Dudarev format was employed~\cite{Dudarev:prb} in the magnetic DFT calculations. Both the lattice constants and atomic positions were fully relaxed with different spin configurations until the Hellman-Feynman force on each atom was smaller than $0.01$ eV/{\AA}. In addition to the standard DFT calculations, the maximally localized Wannier functions (MLWFs) method was employed to fit the Fe $3d$'s bands by using the WANNIER90 packages~\cite{Mostofi:cpc}. All the crystal structures were visualized with the VESTA code~\cite{Momma:vesta}.

\subsection{B. Multi-orbital Hubbard Model}
To better understand the magnetic behavior of the quasi-one-dimensional Ba$_2$FeS$_3$ in the dominant chain direction, an effective multi-orbital Hubbard model was constructed. The model studied here includes the kinetic energy and interaction energy terms $H = H_k + H_{int}$. The tight-binding kinetic portion is described as:
\begin{eqnarray}
H_k = \sum_{\substack{i\sigma\gamma\gamma'}}t_{\gamma\gamma'}(c^{\dagger}_{i\sigma\gamma}c^{\phantom\dagger}_{i+1\sigma\gamma'}+H.c.)+ \sum_{i\gamma\sigma} \Delta_{\gamma} n_{i\gamma\sigma},
\end{eqnarray}
where the first part represents the hopping of an electron from orbital $\gamma$ at site $i$ to orbital $\gamma'$ at the NN site $i+1$, using a chain of length $L$. $\gamma$ and $\gamma'$ represent the three different orbitals.

The (standard) electronic interaction portion of the Hamiltonian is:
\begin{eqnarray}
H_{int}= U\sum_{i\gamma}n_{i\uparrow \gamma} n_{i\downarrow \gamma} +(U'-\frac{J_H}{2})\sum_{\substack{i\\\gamma < \gamma'}} n_{i \gamma} n_{i\gamma'} \nonumber \\
-2J_H  \sum_{\substack{i\\\gamma < \gamma'}} {{\bf S}_{i\gamma}}\cdot{{\bf S}_{i\gamma'}}+J_H  \sum_{\substack{i\\\gamma < \gamma'}} (P^{\dagger}_{i\gamma} P_{i\gamma'}+H.c.).
\end{eqnarray}
The first term is the intraorbital Hubbard repulsion. The second term is the electronic repulsion between electrons at different orbitals where the standard relation $U'=U-2J_H$ is assumed due to rotational invariance. The third term represents the Hund's coupling between electrons occupying the iron $3d$ orbitals. The fourth term is the pair hopping between different orbitals at the same site $i$, where $P_{i\gamma}$=$c_{i \downarrow \gamma} c_{i \uparrow \gamma}$.

To solve the multi-orbital Hubbard model, and obtain the magnetic properties of Ba$_2$FeS$_3$ (HP) along the $b$-axis direction including quantum fluctuations, the many-body technique was employed based on the DMRG method~\cite{white:prl,white:prb}, where specifically we used the DMRG++ software~\cite{Alvarez:cpc}. In our DMRG calculations, we employed a $16$-sites cluster chain with open-boundary conditions (OBC). Furthermore, at least $1400$ states were kept and up to $21$ sweeps were performed during our DMRG calculations. In addition, the electronic filling $n = 4$ in the three orbital was considered. This electronic density (three electrons in four orbitals) is widely used in the context of iron superconductors with Fe$^{\rm 2+}$ valence (n = 6)~\cite{osmp1,Luo:prb10}. The common rationalization to justify this density is to consider one orbital doubly occupied and one empty, and thus both can be discarded. This leads to four electrons in the remaining three orbitals, providing a good description of the physical properties for the real iron systems with $n= 6$~\cite{osmp1,Daghofer:prb10,Rin:prl}.

In the tight-binding term, we used the Wannier function basis \{$d_{xz}$, $d_{x^2-y^2}$, $d_{xy}$\}, here referred to as $\gamma$ =  \{0, 1, 2\}, respectively. We only considered the NN hopping matrix:
\begin{equation}
\begin{split}
t_{\gamma\gamma'} =
\begin{bmatrix}
          0.012     &   0.045  &   0.080	   	       \\
          0.045     &   0.112  &  -0.018	   	       \\
         -0.080	    &   0.018  &   0.238	
\end{bmatrix}.\\
\end{split}
\end{equation}

All the hopping matrix elements are given in eV units. $\Delta_{\gamma}$ is the crystal-field splitting of orbital $\gamma$. Specifically, $\Delta_{0} =-0.339$, $\Delta_{1} = 0.047$, and $\Delta_{2} = -0.127$ (the Fermi level is considered to be zero). The total kinetic energy bandwidth $W$ is 1~eV. More details about the Wannier functions and hoppings can be found in APPENDIX A.

\section{III. DFT results}

\subsection{A. Non-magnetic state}
Before addressing the magnetic properties, let us discuss the electronic structures of the non-magnetic state of Ba$_2$FeS$_3$ (HP) based on the experimentally available structural properties~\cite{Guan:jac}. At high pressure, the lattice constants are $a = 8.683$ ~\AA, $b = 4.297$ \AA~and $c = 17.025$ \AA, respectively.

\begin{figure}
\centering
\includegraphics[width=0.48\textwidth]{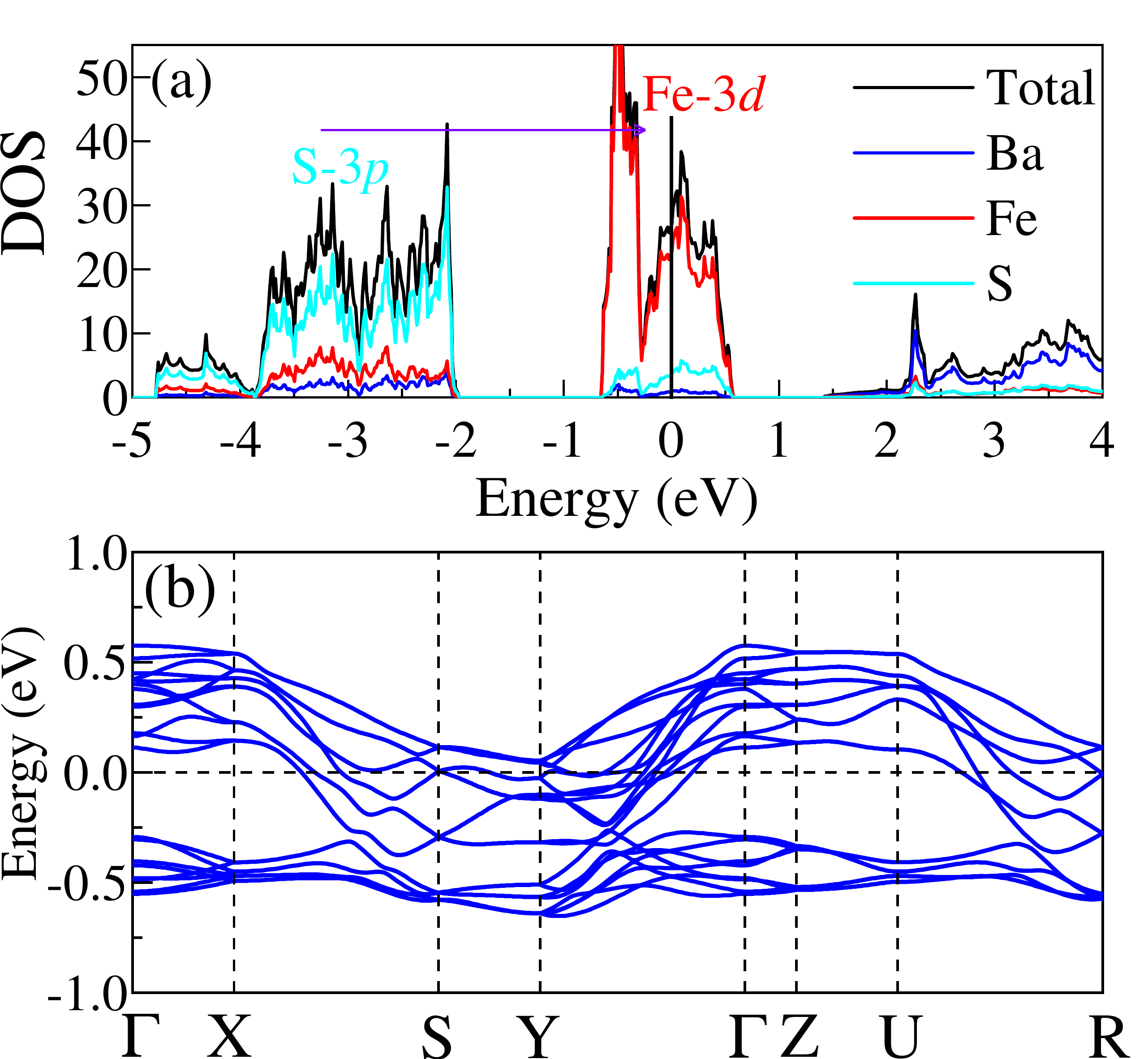}
\caption{ (a) Density-of-states near the Fermi level of Ba$_2$FeS$_3$ (HP) for the non-magnetic phase (black = Total; blue = Ba; red = Fe; cyan = S). (b) Band structures of Ba$_2$FeS$_3$ (HP) for the non-magnetic state. The Fermi level is shown with a dashed horizontal line. The coordinates of the high-symmetry points in the Brillouin zone (BZ) are  $\Gamma$ = (0, 0, 0), X = (0.5, 0, 0), S = (0.5, 0.5, 0), Y = (0, 0.5, 0), Z = (0, 0, 0.5),  U = (0.5, 0, 0.5),  R = (0.5, 0.5, 0.5). Note that all the high-symmetry points are in scaled units, corresponding to the units of $2\pi$/s, ($s = a, ~b$ or $c$).}
\label{Fig2}
\end{figure}

First, we present the density-of-states (DOS) of the non-magnetic state of Ba$_2$FeS$_3$ (HP), displayed in Fig.~\ref{Fig2}(a). Near the Fermi level, the electronic density is mainly contributed by the iron $3d$ orbitals, where the $p-d$ hybridization between Fe $3d$ and S $3p$ states is weak. Furthermore, the Fe $3d$ bands of Ba$_2$FeS$_3$ (HP) are located in a relatively small range of energy from $-1$ to $1$ eV, while the S $3p$ bands are located at a deeper energy level from $-5$ eV to $-2$ eV. In iron ladders~\cite{Zhang:prb19}, the $p-d$ hybridization was reported to be stronger than in the Ba$_2$FeS$_3$ (HP) chain under investigation here. According to the DOS of Ba$_2$FeS$_3$ (HP), the charge transfer gap $\Delta$ = $\varepsilon_d$ - $\varepsilon_p$ is large, indicating Ba$_2$FeS$_3$ (HP) is a Mott-Hubbard system. Thus, the Fe-S hybridization is smaller than that in iron ladders.

The result of the previous paragraph can be understood intuitively. First, in the dominant Fe chain of Ba$_2$FeS$_3$ (HP), the Fe-Fe bond along the chain is about $4.30$ \AA, larger than the corresponding number for the iron ladder ($\sim$ $2.64$ \AA) with $n = 6$~\cite{chi:prl}. Second, there is only one S atom connecting NN Fe atoms (with Fe-S bond $\sim$ $2.44$ ~\AA) in the iron chains of Ba$_2$FeS$_3$ (HP). On the other hand, in iron ladders with $n = 6$~\cite{chi:prl}, there are two S atoms connecting the NN Fe atoms along the leg direction (with Fe-S bonds being $2.29$ and $2.27$~\AA). Considering those differences of structural geometries as compared to iron ladders, the overlap of Fe and S atoms of Ba$_2$FeS$_3$ (HP) would be weaker than that of iron ladders, resulting in a weaker $p-d$ hybridization in the Ba$_2$FeS$_3$ (HP) chain.

The projected band structures of Ba$_2$FeS$_3$ (HP) are displayed in Fig.~\ref{Fig2}(b). It is clearly shown that the band is more dispersive from Y to $\Gamma$ along the chains than along other directions, such as
$\Gamma$ to X along the $a$-axis,
which is compatible with the presence of quasi-one-dimensional chains along the $k_y$ axis. In this case, the intrachain coupling should play the key role in magnetism and other physical properties.

\subsection{B. Magnetism}

\begin{figure}
\centering
\includegraphics[width=0.48\textwidth]{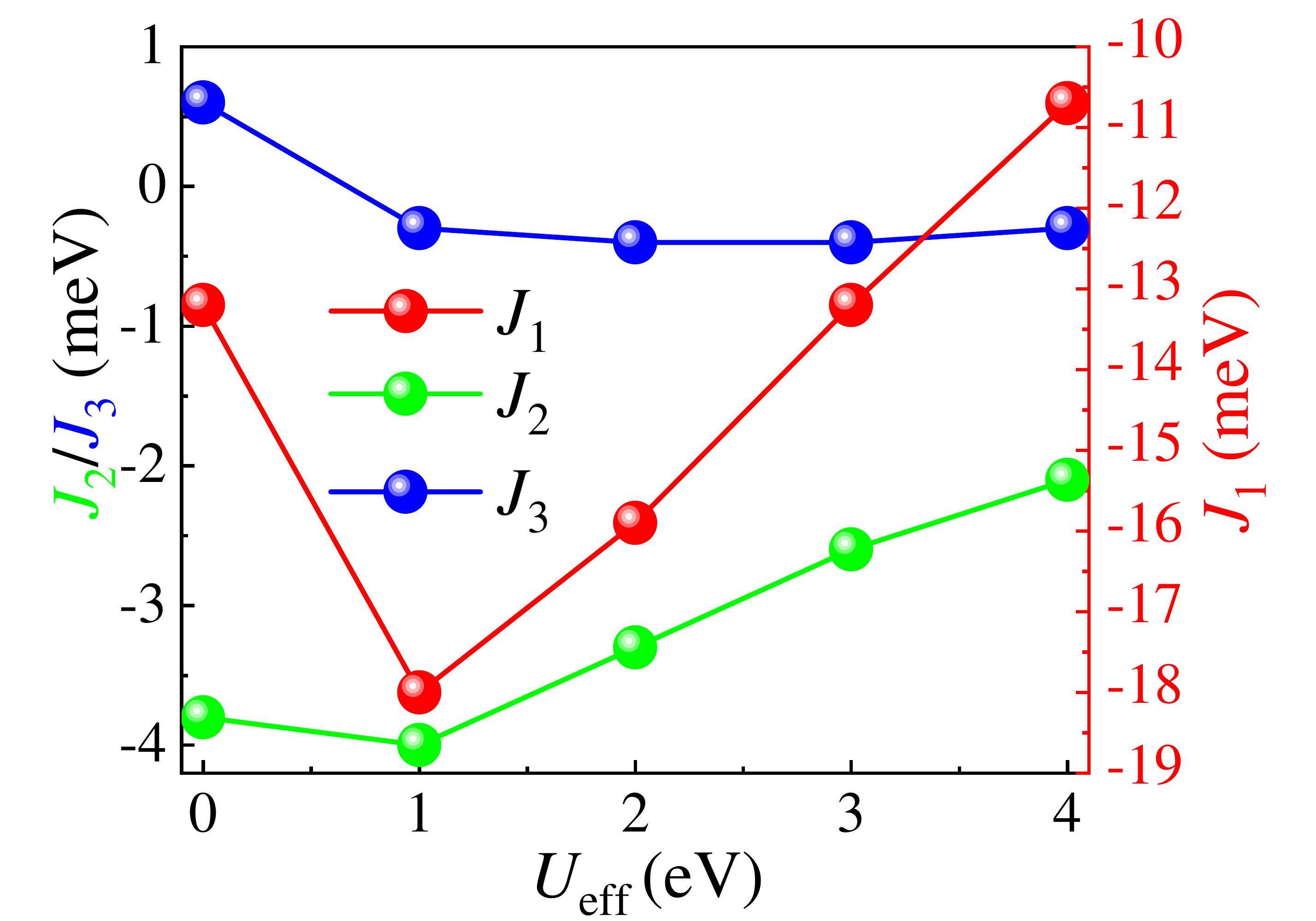}
\caption{Magnetic exchange couplings as a function of the Hubbard $U_{\rm eff}$ coupling.}
\label{Fig3}
\end{figure}

To qualitatively represent the magnetism of Ba$_2$FeS$_3$ (HP), a simple classical Heisenberg model with three magnetic exchange couplings $J$ was introduced to described phenomenologically this system:
\begin{eqnarray}
\nonumber H&=&-J_1\sum_{<ij>}\textbf{S}_i\cdot\textbf{S}_j-J_2\sum_{[kl]}\textbf{S}_k\cdot\textbf{S}_l\\
&&-J_3\sum_{\{mn\}}\textbf{S}_m\cdot\textbf{S}_n,
\end{eqnarray}
where $J_1$ is the intrachain exchange interactions between NN Fe-Fe spin pairs, while $J_2$ and $J_3$ are the interchain exchange interactions between two NN iron chains, corresponding to two different interchain Fe-Fe distances, as displayed in Fig.~\ref{Fig1}(c). By mapping the DFT energies of different magnetic
configurations~\cite{Jcontext}, based on the experimental lattice structure, we obtained the coefficients of different $J$'s as a function of Hubbard $U_{\rm eff}$ in Fig.~\ref{Fig3}. As expected, $J_1$ is the dominant one, indicating that the intrachain magnetic coupling plays the key role in this system. Based on these calculated $J$'s, the magnetic coupling along the chain favours AFM, while the interchain couplings are quite weak.

\begin{figure}
\centering
\includegraphics[width=0.48\textwidth]{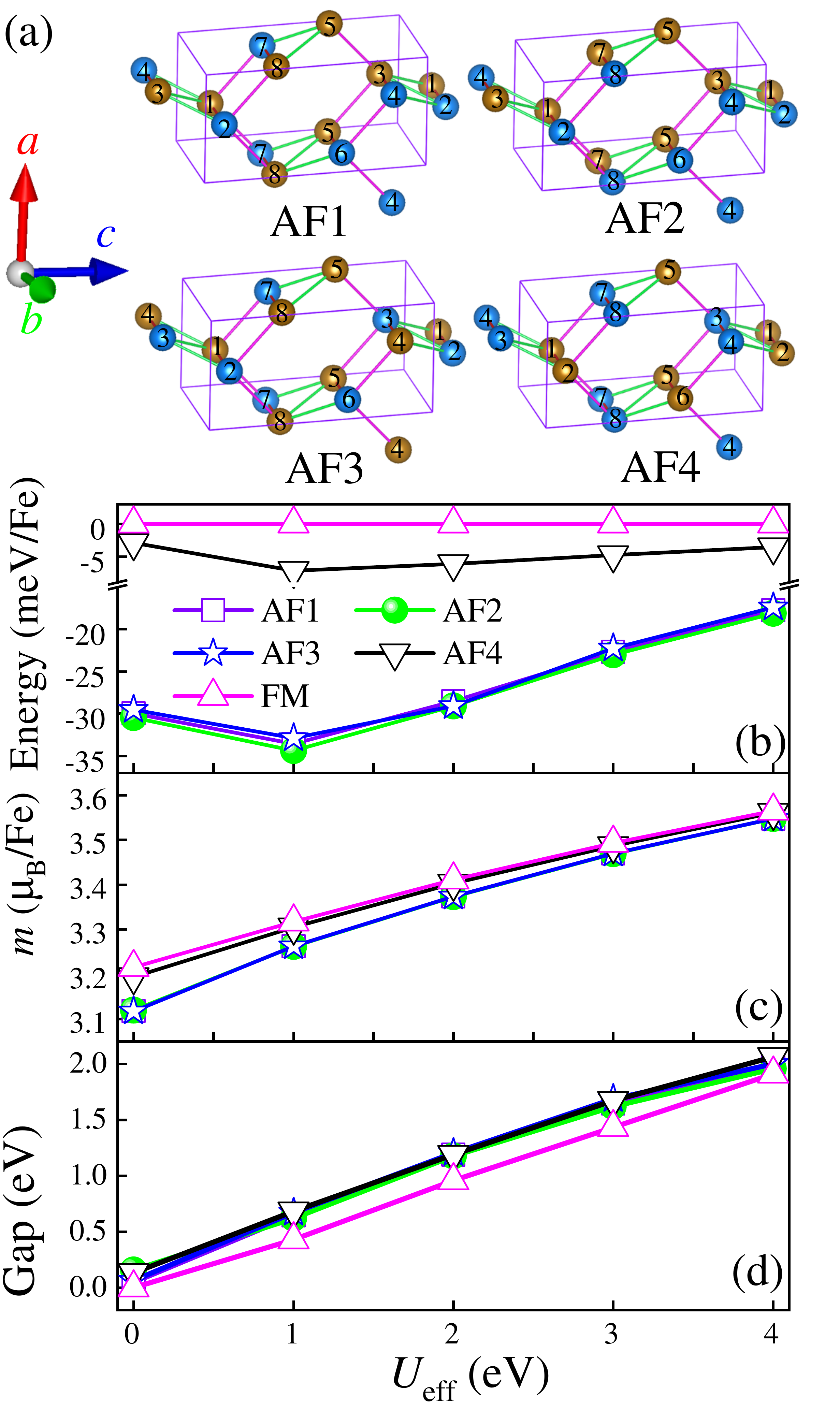}
\caption{(a) Sketch of some possible AFM patterns studied here. Spin up and down are distinguished by brown and blue, respectively. (b-d) DFT results for Ba$_2$FeS$_3$ (HP) as a function of $U_{\rm eff}$. (b) Energies (per Fe) of various magnetic orders are indicated. The FM configuration is taken as the reference. (c) Local magnetic moments of Fe, integrated within the default Wigner-Seitz sphere as specified by VASP. (d) Band gaps for the many states analyzed.}
\label{Fig4}
\end{figure}

To better understand the possible magnetic configurations, we also considered several AFM configurations in a $1\times2\times1$ supercell, as shown in Fig.~\ref{Fig4}(a). Both the lattice constants and atomic positions were fully relaxed for those different spin configurations. First, the AF2 magnetic order always has the lowest energy among all tested candidates, independently of $U_{\rm eff}$, as shown in Fig.~\ref{Fig4}(b). Furthermore, the energies of the AF1 and AF3 orders are close to the energy of the AF1 state, indicating a quite weak $J_3$ coupling, in agreement with our previous discussion using the  Heisenberg model.

Moreover, the calculated local magnetic moment per Fe is displayed in Fig.~\ref{Fig4}(c), for different possible magnetic configurations. With increasing $U_{\rm eff}$, the moment of Fe in the AF2 state increases from $3.12$ to $3.55$ $\mu_{\rm B}$/Fe, which is higher than those calculated for the CX-AFM type configuration in iron ladders with $n=6$~\cite{Zhang:prb17}. As shown in Fig.~\ref{Fig4}(d), all AFM orders are insulating and the gap increases with $U_{\rm eff}$, as expected.

\begin{figure}
\centering
\includegraphics[width=0.48\textwidth]{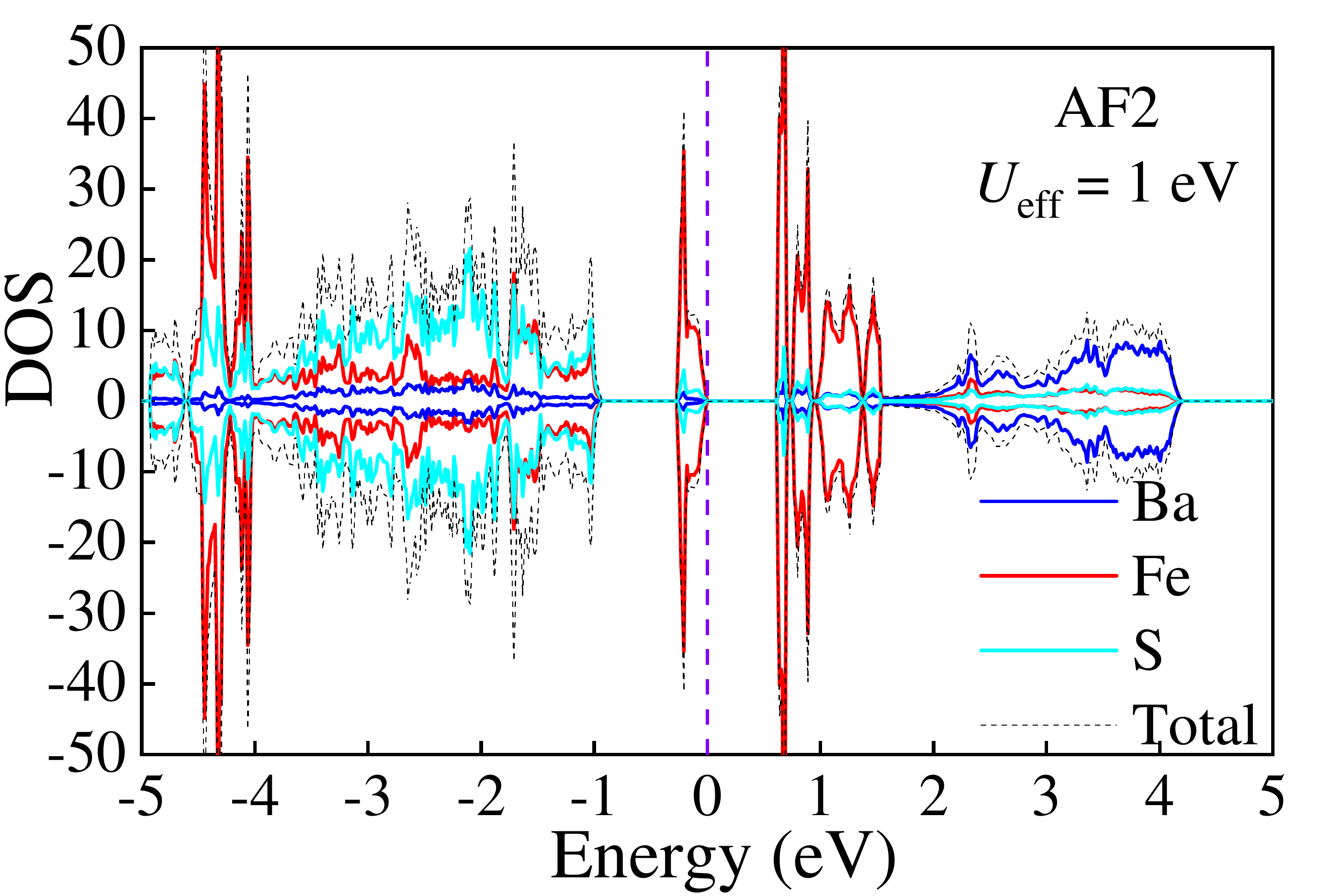}
\caption{DOS for the AF2 state obtained using LSDA + $U_{\rm eff}$ (= 1 eV) DFT caculations.}
\label{Fig5}
\end{figure}

According to the calculated DOS for the AF2 state, the bands near the Fermi level are mostly contributed by Fe's $3d$ orbitals and the Fe atom is in the high spin configuration. Furthermore, the DOS plot indicates a Mott transition behavior [Fig.~\ref{Fig5}]. Our calculated band gap for the case $U_{\rm eff} = 1$ eV in the AF2 state is about $0.62$ eV, which is very close to the experimental gap obtained from fitting the resistivity versus $1/T$ curve ($\sim$ $0.676$ eV)~\cite{Guan:jac}.

In summary of our DFT results, we found a strongly anisotropic quasi-one-dimensional electronic band structure, corresponding to its dominant chain geometry. In addition,we found the AF2 magnetic order is the most likely ground state, where the interchain coupling dominates. Furthermore, our calculations also indicated this system is a Mott Hubbard system with a Mott gap.

\section{IV. DMRG results}
As discussed in the DFT section, the chain direction is the most important for the physical properties of Ba$_2$FeS$_3$ (HP). Using DFT+$U$ calculations, we obtained a strong Mott insulating AFM phase. However, in one-dimensional systems, quantum fluctuations are important at low temperatures. Because DFT neglects fluctuations, here we employed the many-body DMRG method to incorporate the quantum magnetic couplings in the dominant chain, where quantum fluctuations are needed to fully clarify the true magnetic ground state properties. In fact, in previous well-studied iron 1D ladders and chains, those quantum fluctuations were found to be crucial to understand the magnetic properties~\cite{mourigal:prl15,Herbrych:osmp}. It also should be noticed that the DMRG method has proven to be a powerful technique for discussing low-dimensional interacting systems ~\cite{Schollw:rmp05,Stoudenmire:ARCMP}.

As discussed before, here we consider the effective multi-orbital Hubbard model with four electrons in three orbitals per site (more details can be found in Section II-B), corresponding to the electronic density per orbital $n = 4/3$. Note that this electronic density is widely used in the context of iron low dimensional compounds with DMRG technology, where the ``real''  iron is in a valence Fe$^{\rm 2+}$, corresponding to six electrons in five orbitals per site~\cite{osmp1,Luo:prb10}. To understand the physical properties of this system, we measured several observables based on the DMRG calculations.

The spin-spin correlation in real space are defined as
\begin{eqnarray}
S(r)=\langle{{\bf S}_i \cdot {\bf S}_j}\rangle,
\end{eqnarray}
with $r=\left|{i-j}\right|$, and the spin structure factor is
\begin{eqnarray}
S(q)=\frac{1}{L}\sum_{r}e^{-iqr}S(r).
\end{eqnarray}

The site-average occupancy of orbitals is
\begin{eqnarray}
n_{\gamma}=\frac{1}{L}{\langle}n_{i\gamma\sigma}\rangle.
\end{eqnarray}

The orbital-resolved charge fluctuation is defined as:
\begin{eqnarray}
\delta{n_{\gamma}}=\frac{1}{L}\sum_{i}({\langle}n_{\gamma,i}^2\rangle-{\langle}n_{\gamma,i}{\rangle}^2).
\end{eqnarray}

The local spin-squared, averaged over all sites, is
\begin{eqnarray}
\langle S^2 \rangle=\frac{1}{L}\sum_{i}\langle{{\bf S}_i \cdot {\bf S}_i}\rangle.
\end{eqnarray}

As already explained, the hopping amplitudes were obtained from the {\it ab initio} DFT calculations for Ba$_2$FeS$_3$ (HP) (see APPENDIX A for details). Furthermore, based on the spin-spin correlation and spin structure factor, we calculated the phase diagram of the Ba$_2$FeS$_3$ iron chain with increasing $U/W$ at different Hund couplings $J_H$/$U$, using primarily a system size $L$ = $16$. In the following, we will discuss our main DMRG results at $J_H$/$U$ = 0.25 because this robust $J_H$/$U$ value is believed to be physically realistic
for iron-based superconductors~\cite{Luo:prb10}.

\subsection{A. Staggered AFM phase}
\begin{figure}
\centering
\includegraphics[width=0.48\textwidth]{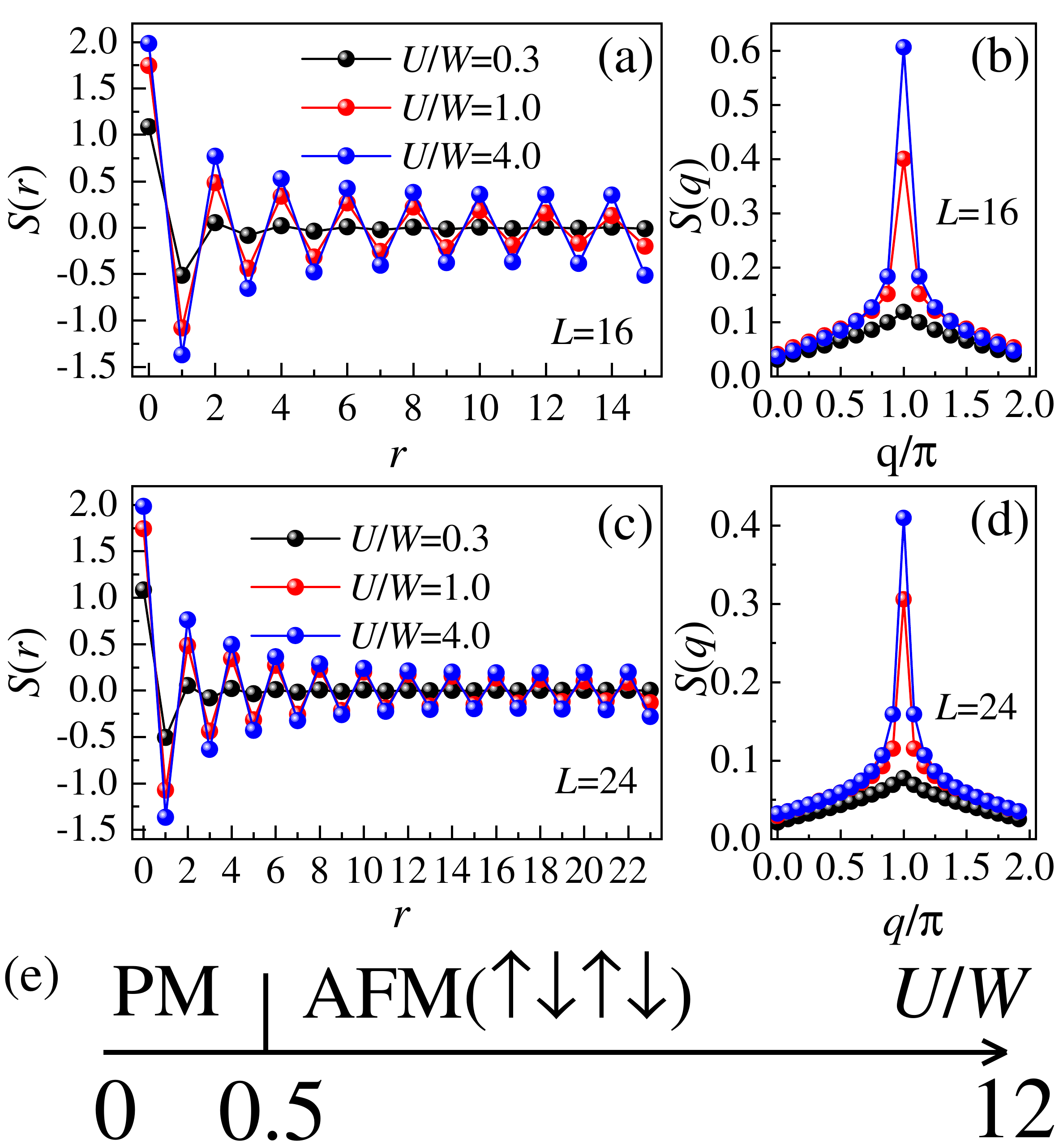}
\caption{(a) Spin-spin correlation $S(r)=\langle{{\bf S}_i \cdot {\bf S}_j}\rangle$ (with $r=\left|{i-j}\right|$ in real space) and (b) the spin structure factor $S(q)$, both at different values of $U/W$, and all at $J_H$/$U$ = 0.25. We use a chain with $L = 16$. (c) Spin-spin correlation $S(r)=\langle{{\bf S}_i \cdot {\bf S}_j}\rangle$ (with $r = \left|{i-j}\right|$ in real space) and (d) the spin structure factor $S(q)$, at different values of $U/W$ and fixed $J_H$/$U$ = 0.25, for $L = 24$. (e) Magnetic phase diagram for $J_H$/$U$ = 0.25.}
\label{Fig6}
\end{figure}

Based on the DMRG measurements of the spin-spin correlation and spin structure factor, we found the paramagnetic phase (PM) at small $U/W$, followed by a robust canonical staggered AFM phase with $\uparrow$-$\downarrow$-$\uparrow$-$\downarrow$ configuration.
Figure~\ref{Fig6}(a) shows the spin-spin correlation $S(r)$=$\langle{{\bf S}_i \cdot {\bf S}_j}\rangle$ vs distance $r$, for different values of $U/W$. The distance is defined as $r=\left|{i-j}\right|$, with $i$ and $j$ site indexes. At small Hubbard interaction $U/W$ $\textless$ 0.6, the spin correlation $S(r)$ decays rapidly vs. distance $r$, indicating paramagnetic behavior, as shown in Fig.~\ref{Fig6}(a) (see result at $U/W$ = 0.3). By increasing $U/W$, the system transitions to the canonical staggered AFM phase with the $\uparrow$-$\downarrow$-$\uparrow$-$\downarrow$ configuration in the whole region of our study ($U/W \le 12$). As shown in Fig.~\ref{Fig6}(b), the spin structure factor $S(q)$ displays a sharp peak at $q = \pi$, corresponding to the canonical staggered AFM phase, consistent with our DFT calculations. In addition, we also calculated the spin-spin correlation $S(r)$ and spin structure factor $S(q)$ using a larger cluster $L = 24$, as shown in Figs.~\ref{Fig6}(c-d). Those results are similar to the results of $L = 16$, indicating that our conclusions of having a canonical staggered AFM phase with $\pi$ vector dominating in the phase diagram  is robust against changes in $L$. Note that in one dimension,
quantum fluctuations prevent full long-range order. Thus, the tail of the spin-spin correlations have a smaller value for $L=24$ than for $L=16$. But the staggered order tendency is clear in both cases.

In the range of $U/W$ we studied, we did not observe any other magnetic ordering tendencies, suggesting the AFM coupling($\uparrow$-$\downarrow$-$\uparrow$-$\downarrow$) is quite stable. This is physically reasonable, considering known facts about the Hubbard model. In the Ba$_2$FeS$_3$ (HP) system, the iron $3d$ orbitals are mainly located in a small energy region and with small bandwidth ($\sim 1$ eV), as shown in Fig.~\ref{Fig2}. By introducing the on-site Hubbard $U$ interaction on Fe sites, the $3d$ orbitals would be easily localized in the Fe sites because the bandwidth is narrow. In this case, the standard superexchange Hubbard spin-spin interaction dominates, leading the spins to order antiferromagnetically along the chain. Note that one orbital ($\gamma=2$) clearly has the largest hopping amplitude from the DFT results, thus this orbital leads in the formation of the AFM order. Due to the large Fe-Fe distance  ($\sim 4.3$ \AA) and the special FeS$_4$ chain geometry, in the Ba$_2$FeS$_3$ (HP) system the electrons of the iron $3d$ states are localized with weak $p-d$ hybridization, dominating the superexchange mechanism. Hence, our DMRG results indicating the dominance of the $\uparrow$-$\downarrow$-$\uparrow$-$\downarrow$ configuration are in agreement with our DFT calculations.

\subsection{B. Charge fluctuations}

\begin{figure}
\centering
\includegraphics[width=0.48\textwidth]{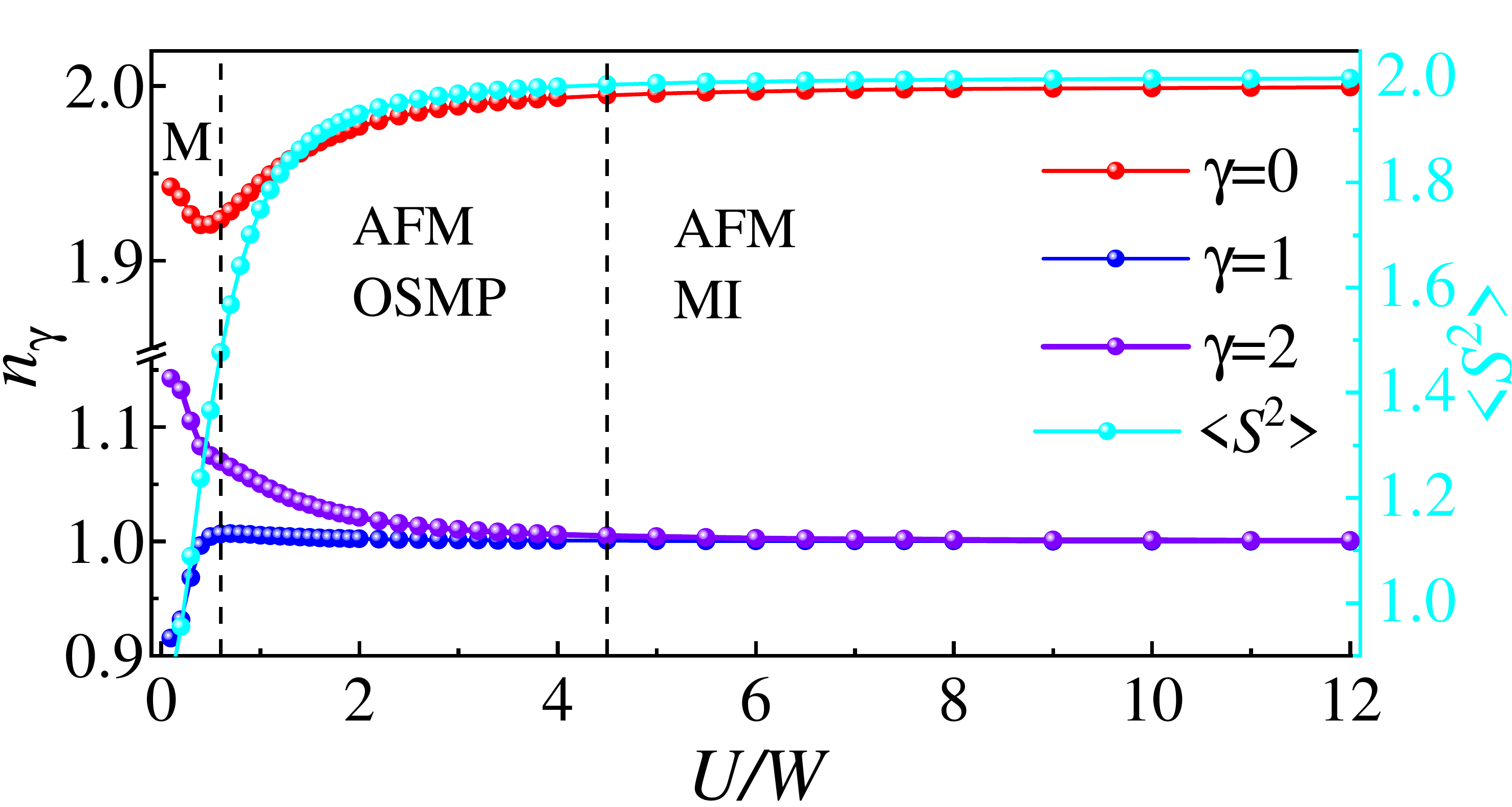}
\caption{Orbital-resolved occupation number $n_{\gamma}$, averaged value of the total spin-squared $\langle{{S}}^2\rangle$ vs. $U/W$, at $J_{H}/U = 1/4$. Here, we used a $16$-sites cluster chain with NN hoppings for four electrons in three orbitals.}
\label{Fig7}
\end{figure}

The site-average occupancy of different orbitals $n_{\gamma}$ vs $U/W$ is shown in Fig.~\ref{Fig7}, for a typical value of $J_H/U$. At small $U/W$ ($\textless 0.6$), a metallic weakly-interacting state is found, with non-integer $n_{\gamma}$ values. In the other extreme of much larger $U/W$, the population of orbital $\gamma = 0$ reaches $2$, and this orbital decouples from the system. Furthermore, the other two orbitals $\gamma = 1$ and $\gamma = 2$ reach population $1$, leading to two half-occupied states. In this extreme $U/W$ case ($n_0$ = 2, $n_1$ =1 and $n_2$ = 1), the system is in a Mott insulator staggered AFM state.

In addition, the average value of the local spin-squared averaged over all sites $\langle{{S}}^2\rangle$ is also displayed in Fig.~\ref{Fig7}, varying $U/W$. The strong local magnetic moments are fully developed with spin magnitude S $\sim 1$, corresponding to four electrons in three orbitals at very large $U/W$.

\begin{figure}
\centering
\includegraphics[width=0.48\textwidth]{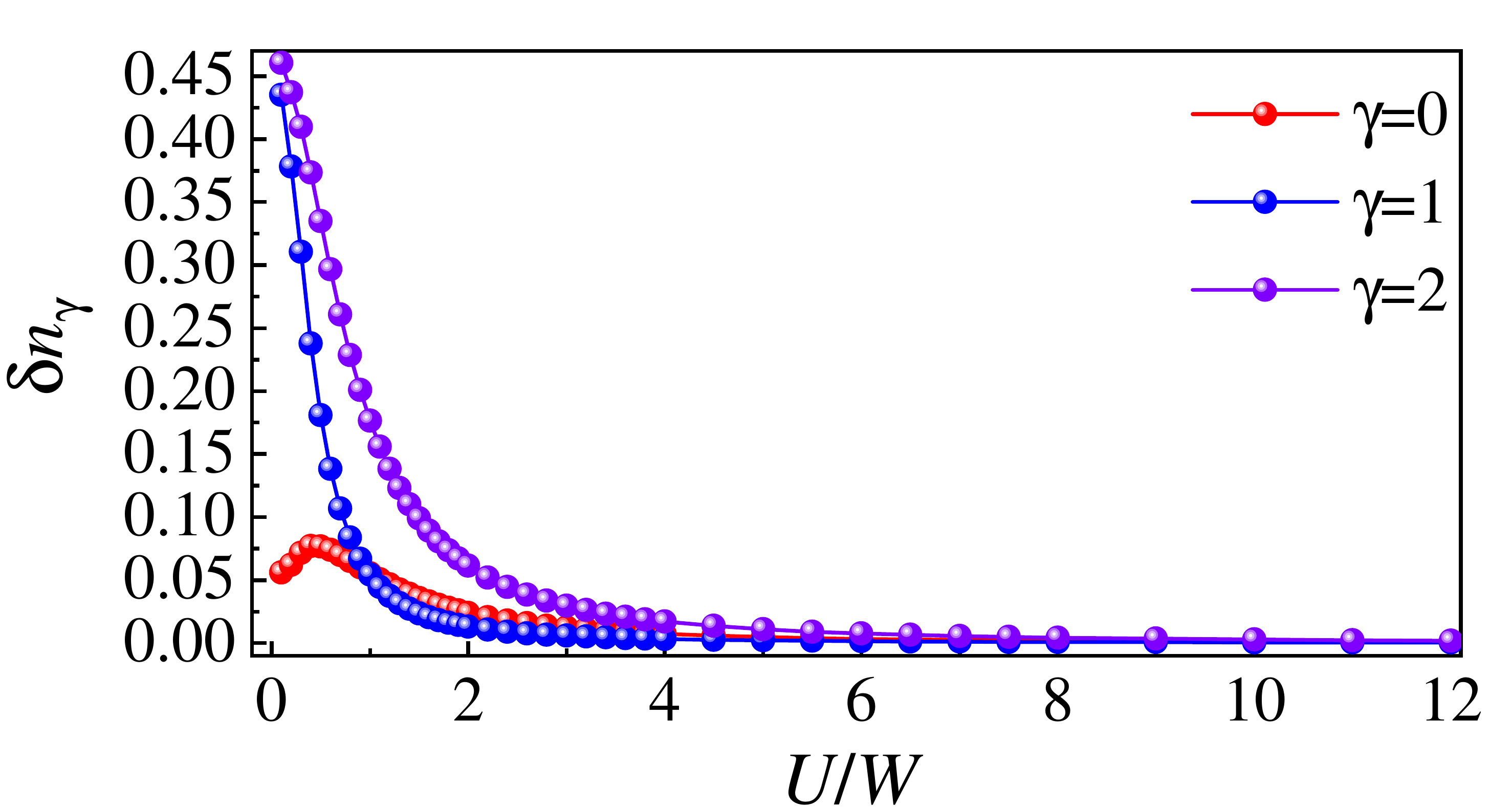}
\caption{Charge fluctuations $\delta{n_{\gamma}}=\frac{1}{L}\sum_{i}({\langle}n_{\gamma,i}^2\rangle-{\langle}n_{\gamma,i}{\rangle}^2)$ vs $U/W$ at $J_{H}/U = 1/4$. Here, we used a $16$-sites cluster chain with NN hoppings for four electrons in three orbitals.}
\label{Fig8}
\end{figure}

To better understand the characteristics of metallic vs insulating behavior in this system, we have also studied the charge fluctuations $\delta{n_{\gamma}}$ for different orbitals, as shown in Fig.~\ref{Fig8}. In the small-$U$ paramagnetic phase ($U/W \textless 0.6$), the system is metallic due to weak interactions. Increasing $U/W$, the charge fluctuations of different orbitals are considerable at intermediate Hubbard coupling strengths, indicating strong quantum fluctuations along the chains. As $U/W$ increases further, the charge fluctuations of $\gamma = 1$ rapidly reaches zero, leading to localized orbital characteristics, while the $\gamma = 2$ orbital still has larger fluctuations with some itinerant electrons. In this case, this intermediate regime corresponds to the OSMP state. At even larger $U/W$ ($\gtrsim 4.5$), the charge fluctuations of the different orbitals are suppressed to nearly zero. Thus, the system becomes fully insulating at very large $U/W$, with two half-filled orbitals ($\gamma = 1$ and $\gamma = 2$) and one fully occupied orbital ($\gamma = 0$) [see Fig.~\ref{Fig7}], as already explained. Here, the charge fluctuations are totally suppressed by the electronic correlations.

\subsection{C. Orbital-selective Mott phase}

Let us now focus on the intermediate regime of OSMP. As displayed in Fig.~\ref{Fig7}, at intermediate Hubbard coupling strengths, the system displays OSMP behavior. In this regime, the $\gamma = 1$ orbital population reaches $1$, indicating localized electronic characteristics, while the other two orbitals have non-integer electronic density, leading to metallic electronic features. Furthermore, we also compare these results with a larger system site $L = 24$ (see APPENDIX B), indicating the conclusion is robust against changes in $L$. Although the site-average occupancy is $1$ (see Fig.~\ref{Fig7}), the $\gamma = 1$ orbital has some charge fluctuations in the region $0.6 \lesssim U/W \textless 2.0$, as shown in Fig.~\ref{Fig8}. Above $U/W = 2.0$, the charge fluctuations of the $\gamma = 1$ orbital remain zero, indicating full localized behavior, while the other two orbitals still have finite values for the charge fluctuations until a larger $U/W \sim 5$.

\begin{figure}
\centering
\includegraphics[width=0.48\textwidth]{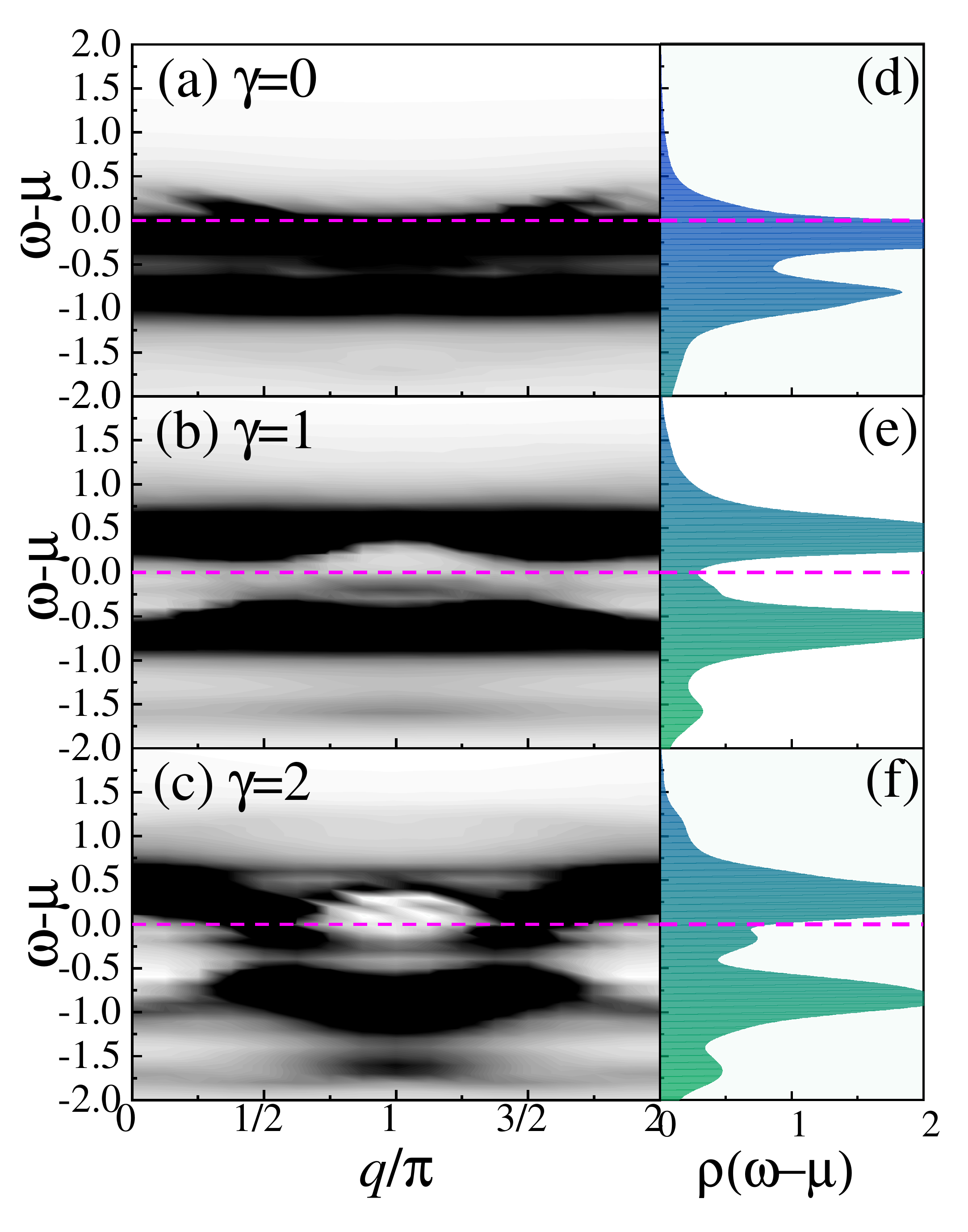}
\caption{(a-c) Single-particle spectra $A_{\gamma}$($q, \omega$) for different orbitals at $U/W$ = 1 and $J_H/U = 0.25$. (d-f) Orbital-resolved PDOS} $\rho_{\gamma}$($\omega$) for different orbitals at $U/W = 1.0$ and $J_H/U = 0.25$.
\label{Fig9}
\end{figure}

To better understand the OSMP region, we calculated the single-particle spectra $A_{\gamma}$($q, \omega$) and the orbital-resolved projected  density of states (PDOS) $\rho_{\gamma}(\omega)$ vs. frequency $\omega$ by using the dynamical DMRG, where the dynamical correlation vectors were obtained using the Krylov-space approach~\cite{Kuhner:prb,Nocera:pre}. Here, the broadening parameter $\eta = 0.1$ was chosen in our DMRG calculations. The chemical potential is obtained from $\mu = (E_{N+1} - E_{N-1})/2$, where $E_N$ is the ground state energy of the $N$-particle system. The single-particle spectra $A_{\gamma}(q, \omega)= A_{\gamma}(q, \omega \textless \mu)+ A_{\gamma}(q, \omega \textgreater \mu)$ is calculated from the portions of the spectra below and above $\mu$, respectively.
\begin{eqnarray}
&&\nonumber A_{\gamma}(q, \omega \textless \mu)=\frac{-1}{L\pi} \sum_{j}e^{ijq}\\
&& Im \left< \Psi_{\rm GS} \left|c_{j,\gamma} \frac{1}{\omega -H + E_G+i\eta}c_{L/2,\gamma} \right|\Psi_{\rm GS}\right>,
\end{eqnarray}
\begin{eqnarray}
&&\nonumber A_{\gamma}(q, \omega \textgreater \mu)=\frac{-1}{L\pi} \sum_{j}e^{ijq}\\
&& Im \left< \Psi_{\rm GS} \left| c_{j,\gamma}^{\dagger} \frac{1}{\omega +H -E_G+i\eta}c_{L/2,\gamma}^{\dagger} \right|\Psi_{\rm GS} \right>,
\end{eqnarray}
where $j$ is a site, $c_{j,\gamma} = \sum_{\sigma}c_{j,\gamma,\sigma}$ is the fermionic anihilation operator. while $c^{\dagger}_{i,\gamma} = \sum_{\sigma}c^{\dagger}_{j,\gamma,\sigma}$ is the creation operator, $E_G$ is the ground state energy, and $\Psi_{\rm GS}$ is the ground-state wave function of the system.

The corresponding orbital-resolved PDOS $\rho_{\gamma}(\omega)$ was defined as:
\begin{eqnarray}
\rho_{\gamma}(\omega ) = \frac{-1}{\pi}\sum_{q}Im A_{\gamma}(q, \omega).
\end{eqnarray}
where $A_{\gamma}$($q, \omega$) is a single-particle Green's function of the ${\gamma}$ orbital electrons.

\begin{figure}
\centering
\includegraphics[width=0.48\textwidth]{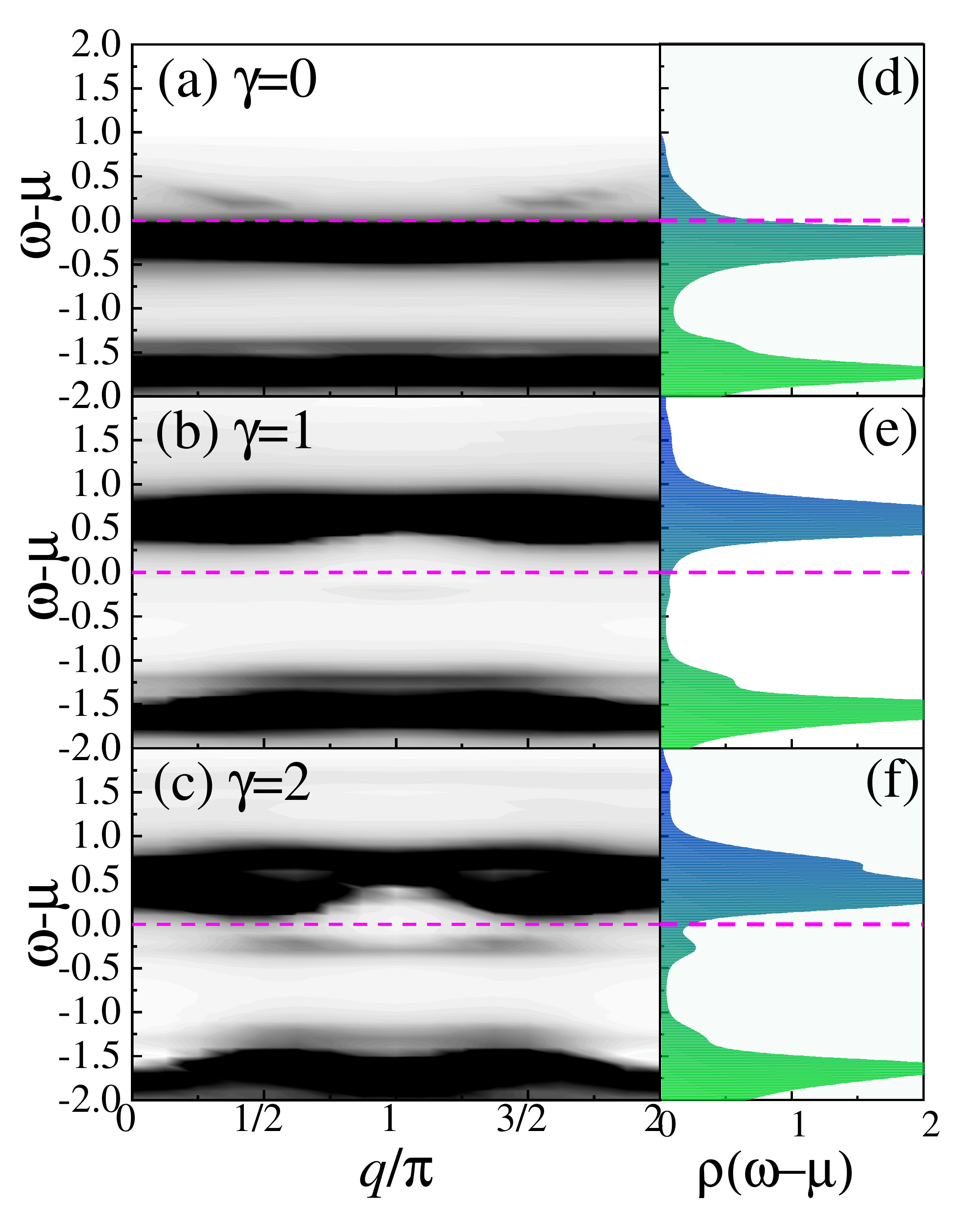}
\caption{(a-c) Single-particle spectra $A_{\gamma}$($q, \omega$) for different orbitals at $U/W$ = 1 and $J_H/U = 0.25$. (d-f) Orbital-resolved PDOS $\rho_{\gamma}$($\omega$) for different orbitals, at $U/W = 2$ and $J_H/U = 0.25$.}
\label{Fig10}
\end{figure}

We calculated the single-particle spectra $A_{\gamma}$($q, \omega$) and PDOS $\rho_{\gamma}(\omega)$ at $J_{H}/U = 0.25$ and $U/W = 1.0$, as displayed in Fig.~\ref{Fig9}. The ${\gamma} = 0$ and ${\gamma} = 2$ orbitals present a metallic behavior, suggesting the electrons are itinerant. Meanwhile, the ${\gamma} = 1$ orbital displays the Mott transition behavior with the pseudogap characteristic, where there are still some finite charge fluctuations in this orbital. In addition, we also present the single-particle spectra $A_{\gamma}$($q, \omega$) and PDOS $\rho_{\gamma}(\omega)$ for $U/W = 2$ in Fig.~\ref{Fig10}. It is clearly shown that the ${\gamma} = 1$ orbital has a Mott gap, while the other two orbitals have some electronic bands crossing the Fermi level, indicating itinerant electronic behavior.

Hence, in this regime of intermediate Hubbard coupling strengths, the coexistence of localized and itinerant carriers supports the OSMP picture. This OSMP is related to having special conditions in the system, such as different bandwidth and crystal fields, as well as intermediate electronic correlation. Here, the ${\gamma} = 1$ orbital is easier to be localized by Hubbard $U$ than the ${\gamma} = 2$ orbital due to different bandwidths. The OSMP physics has been extensively discussed in experimental and theoretical works in low-dimensional iron systems with electronic density $n = 6$, such as the iron ladders BaFe$_2$Se$_3$~\cite{mourigal:prl15,osmp1,osmp2} and the iron pnictides/chalcogenides superconductors~\cite{OSMP,Yi:prl}. Here, our DMRG results indicate this interesting OSMP state may also appear in the Ba$_2$FeS$_3$ iron chain system (electronic density $n = 6$), and they thus deserve more experimental studies. As the electronic correlation $U/W$ increases, all the orbitals eventually become Mott-localized with the electronic occupancies ($n_0$ = 2, $n_1$ =1 and $n_2$ = 1), as displayed in  Fig.~\ref{Fig7}. Then, the MI phase eventually suppresses the OSMP at very large Hubbard coupling.

\begin{figure}
\centering
\includegraphics[width=0.48\textwidth]{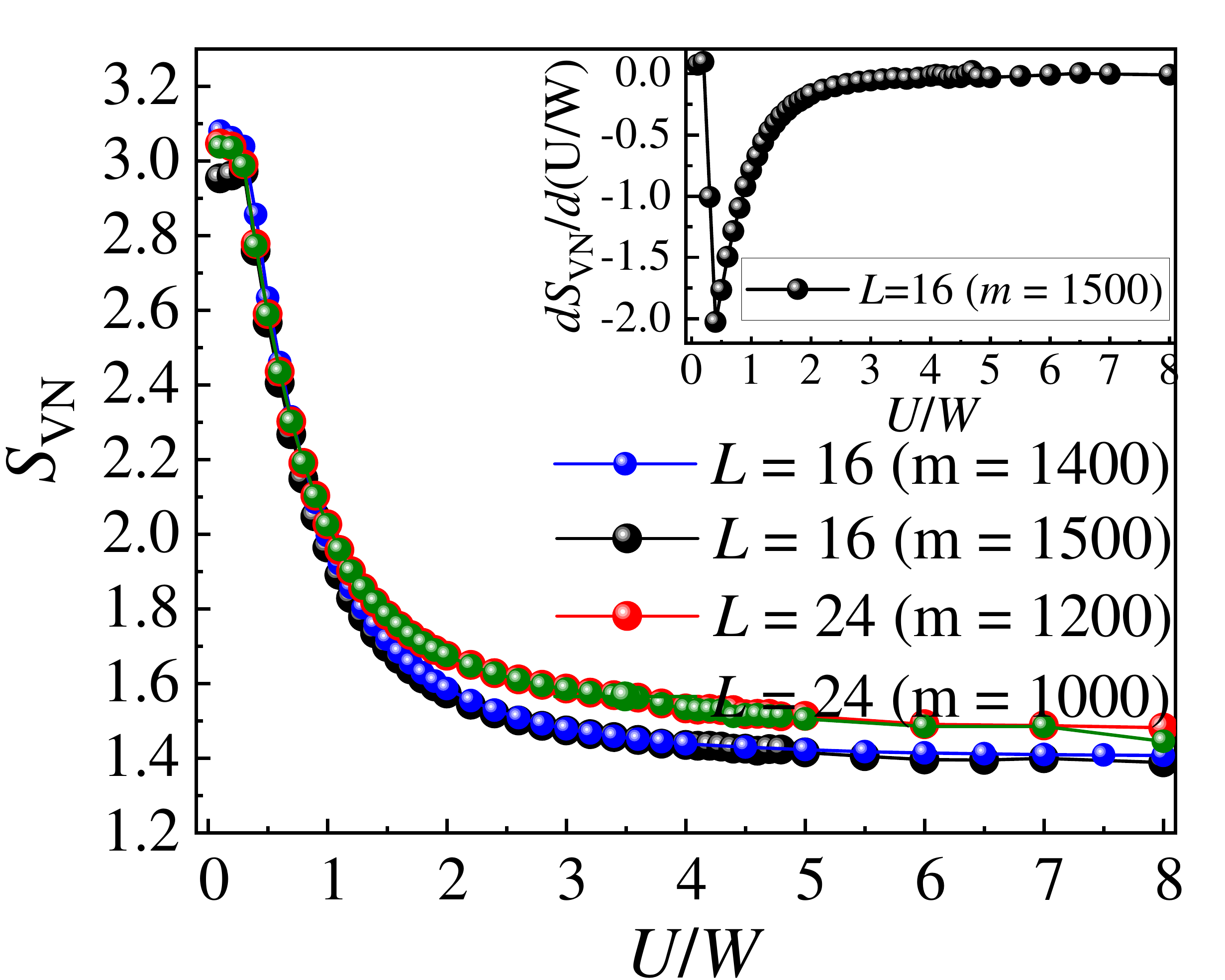}
\caption{Von Neumann entanglement entropy ($S_{\rm VN}$) for the three-orbital chain model as a function of $U/W$ at $J_H$/$U$ = 0.25. inset: derivative of $S_{\rm VN}$}
\label{Fig11}
\end{figure}

In addition, we also calculated the entanglement entropy to better understand the OSMP-MI phase transition, using the Von Neumann form~\cite{Calabrese:JSM,Eisert:rmp}. As shown in Fig.~\ref{Fig11}, there are three regimes here, corresponding to PM, AFM-OSMP, and AFM-MI states, which is qualitatively in agreement with our results via the spin-spin correlation $S(r)$ and charge fluctuations $\delta{n_{\gamma}}$. At $U/W \geq 0.4$, $S_{\rm VN}$ begins to drop rapidly, corresponding to the PM to AFM-OSMP phase transition. At $U/W \geq 4.5$, $S_{\rm VN}$ smoothly converges to a constant. In fact, this convergence does not reflect on the spin-spin correlation $S(r)$ because the magnetic order does not change from the AFM-OSMP state to the AFM-MI phase. The main difference between AFM-MI and AFM-OSMP relies on the electronic density i.e. whether is localized or not. In this case, this difference between those two states can be reflected in the charge fluctuations $\delta{n_{\gamma}}$, where all the orbitals eventually with increasing $U/W$ become Mott-localized leading to insulating behavior starting approximately at $U/W \sim 4.5$ (see Fig.~\ref{Fig7}). It also should be noticed that finite lattice size effects and the use of a limited number of states in DMRG would affect the specific boundary values of this regime change from delocalized to localized electrons. But the presence of three different regimes in this model was established via the entanglement entropy, qualitatively agreeing with our other DMRG results. Since the two states (AFM-OSMP and AFM-MI) involved in the discussion are both AFM, we believe that the transition from OSMP to MI is not a sharp true phase transition involving a singularity in some quantity (see Fig.~\ref{Fig11}). Hence, we believe it can be better described as a ``rapid crossover'' from AFM-OSMP to AFM-MI.

\subsection{D. Additional Results}

\begin{figure}
\centering
\includegraphics[width=0.48\textwidth]{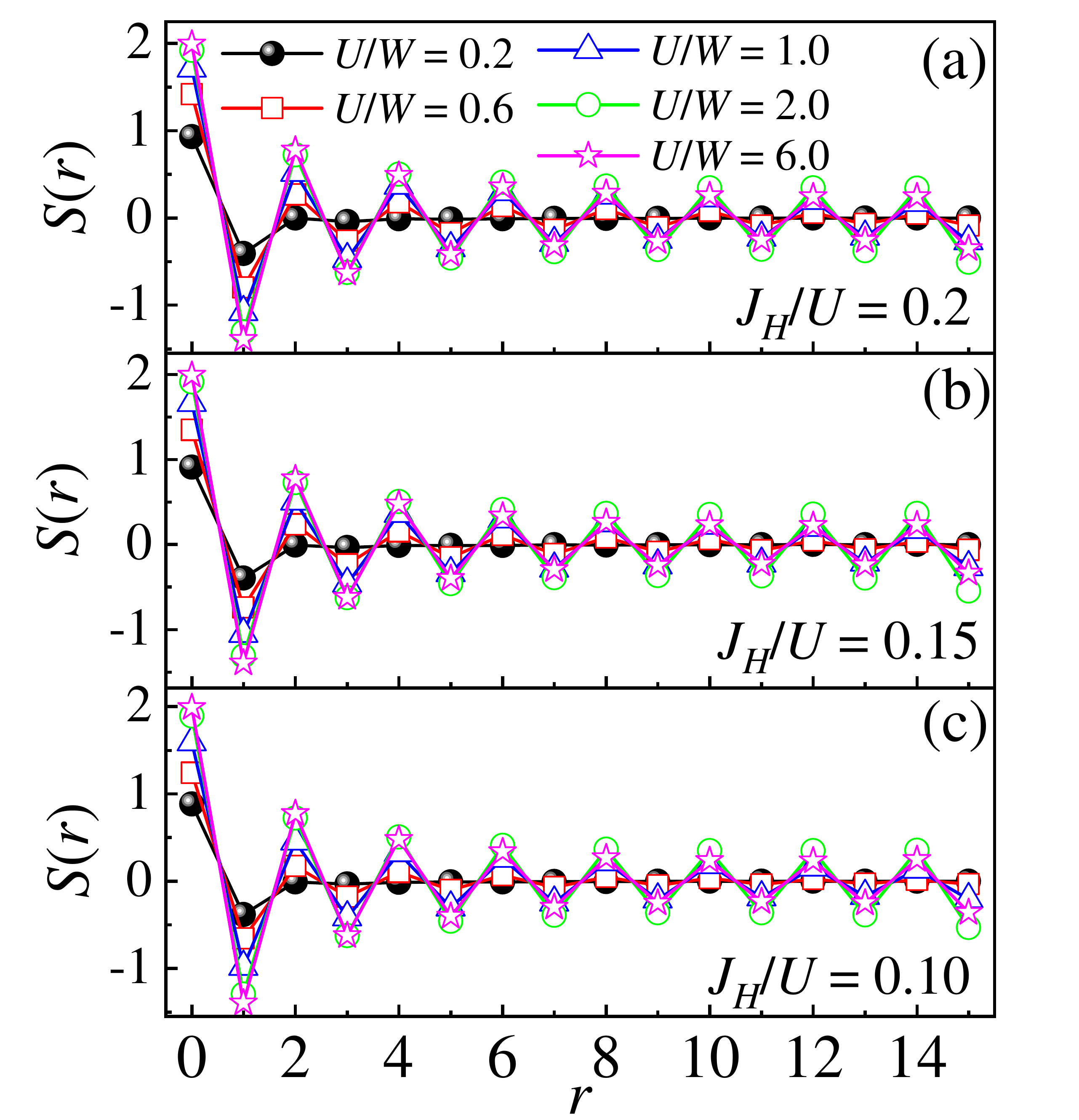}
\caption{Spin-spin correlation $S(r)=\langle{{\bf S}_i \cdot {\bf S}_j}\rangle$ (with $r=\left|{i-j}\right|$) in real space for different values of $U/W$ at (a) $J_H$/$U$ = 0.20, (b) $J_H$/$U$ = 0.15, and (c) $J_H$/$U$ = 0.10.}
\label{Fig12}
\end{figure}

As shown in Fig.~\ref{Fig12}, we present the spin-spin correlation $S(r)$ for several values of $U/W$, at different Hund couplings $J_H$/$U$ = $0.10$, $0.15$, and $0.20$. As the electronic correlation $U/W$ increases, the staggered AFM phase with $\pi$ vector becomes dominant in the entire region, at least within the range we studied. In fact, this staggered AFM order ($\pi$ vector) was also observed in a large regime of the phase diagram in previous mean-field calculations~\cite{Luo:prb14}, although by using different hoppings.

Due to its unique geometric chain configuration, this system displays strong Hubbard superexchange interaction along the chain, which is different from other iron-based chains or ladders. It should be noted that several interesting phases (i.e. block-type $\uparrow$-$\uparrow$-$\downarrow$-$\downarrow$ and FM phases) were found in our previous DMRG phase diagram for a chain system~\cite{Rin:prl,Pandey:prb,Lin:osmp}. Previous work~\cite{Rin:prl} suggests the block-type AFM could be stable due to the competition between the $J_H$ and superexchange interaction. The $J_H$ favors FM ordering, corresponding to the double-exchange interaction in manganites~\cite{Dagotto:rp}, while the superexchange interaction favors AFM ordering. However, in the Ba$_2$FeS$_3$ system of our focus here, the weak $p-d$ hybridization suppresses the double exchange interaction. Thus, the superexchange Hubbard interaction is dominant, leading to robust staggered AFM order. Again, we believe this is because only one of the orbitals has a robust intraorbital hopping, thus dominating the physics.

\begin{figure}
\centering
\includegraphics[width=0.48\textwidth]{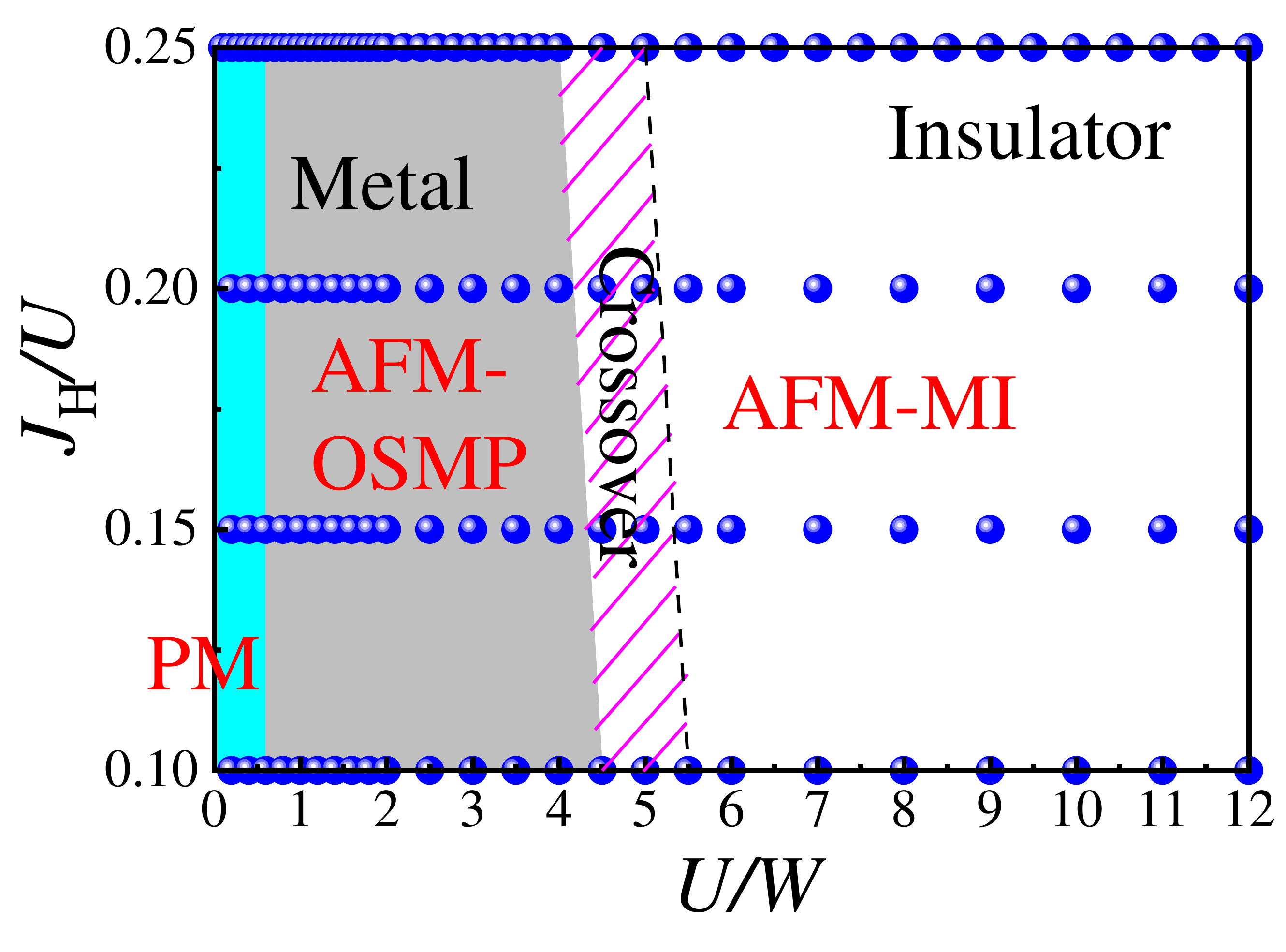}
\caption{DMRG phase diagram of the three-orbital Hubbard model varying $U/W$ and $J_H/U$, using a $L = 16$ chain. Different phases are indicated, with the PM, AFM-OSMP, AFM-MI phasese. Small solid circles indicate specific values of data points that were investigated
with DMRG calculations.}
\label{Fig13}
\end{figure}

In addition, the magnetic phase diagram was calculated varying $J_H$/$U$ and $U/W$, based on the DMRG results (spin-spin correlation $S(r)$ and charge fluctuations $\delta{n_{\gamma}}$). We found three dominant regimes, involving metallic PM, AFM-OSMP, and AFM-MI phases, as shown in Fig.~\ref{Fig13}. Note that the boundaries coupling values should be considered only as crude approximations. However, the existence of the three regions shown was clearly established, even if the boundaries are only rough estimations. We believe our theoretical phase diagram should encourage a more detailed experimental study of iron chalcogenide compounds or related systems.

If the NN distance could be reduced by considering chemical doping or strain effects, the crystal-field splitting and the hybridization would increase. Then, it may be possible to achieve some interesting magnetic phases in this system, as discussed in Refs.{~\cite{Luo:prb14,Rin:prl}}. This maybe a possible direction for further experimental or theoretical studies working on this material or similar variations obtained by altering the 213 chemical formula.

In summary of our DMRG results, after the paramagnetic regime of weak coupling we found the AFM state with $\uparrow$-$\downarrow$-$\uparrow$-$\downarrow$ configuration in our three-orbital Hubbard model, at the robust range of $U/W$ and $J_H$/$U$ that we studied. At intermediate Hubbard coupling strengths, this system displayed OSMP behavior, while the OSMP was suppressed by MI phase at very large $U/W$.

\section{V. Conclusions}

In this publication, we systematically studied the compound Ba$_2$FeS$_3$ (HP) by using first-principles DFT and also DMRG calculations. A strongly anisotropic one-dimensional electronic band structure was observed in the non-magnetic phase, corresponding to its dominant chain geometry. The magnetic coupling along the chain was found to be the key ingredient for magnetism. The staggered magnetic state with a Mott gap was found to be the most likely magnetic ground state among all the candidates studied. Based on the Wannier functions calculated from DFT, we obtained the nearest-neighbor hopping amplitudes and on-site energies for the iron atoms. Then, a multi-orbital Hubbard model for the iron chain was constructed and studied by using the many-body DMRG methodology, considering quantum fluctuations. Based on the DMRG calculations, we obtained a dominant staggered AFM state ($\uparrow$-$\downarrow$-$\uparrow$-$\downarrow$). This staggered $\uparrow$-$\downarrow$ AFM with $\pi$ vector was found in a robust portion of the phase diagram at many values of  $U/W$ and $J_H/U$, in agreement with DFT calculations. At intermediate Hubbard coupling strengths, this system displayed orbital-selective Mott phase behavior, corresponding to one localized orbital and two itinerant metallic orbitals, the latter with nonzero charge fluctuations. At larger $U/W$, the system crossovers to a Mott insulating state ($n_0$ = 2, $n_1$ =1 and $n_2$ = 1) with one double occupied orbital ($\gamma = 0$) and two half occupied orbitals ($\gamma = 1$ and $\gamma = 2$).

\section{Acknowledgments}
The work of Y.Z., L.-F.L., A.M. and E.D. is supported by the U.S. Department of Energy (DOE), Office of Science, Basic Energy Sciences (BES), Materials Sciences and Engineering Division. G.A. was partially supported by the scientific Discovery through Advanced Computing (SciDAC) program funded by U.S. DOE, Office of Science, Advanced Scientific Computing Research and BES, Division of Materials Sciences and Engineering. The calculations were carried out at the Advanced Computing Facility (ACF) of the University of Tennessee, Knoxville.

\section{APPENDIX}
\subsection{A. Hoppings}
\begin{figure}
\centering
\includegraphics[width=0.48\textwidth]{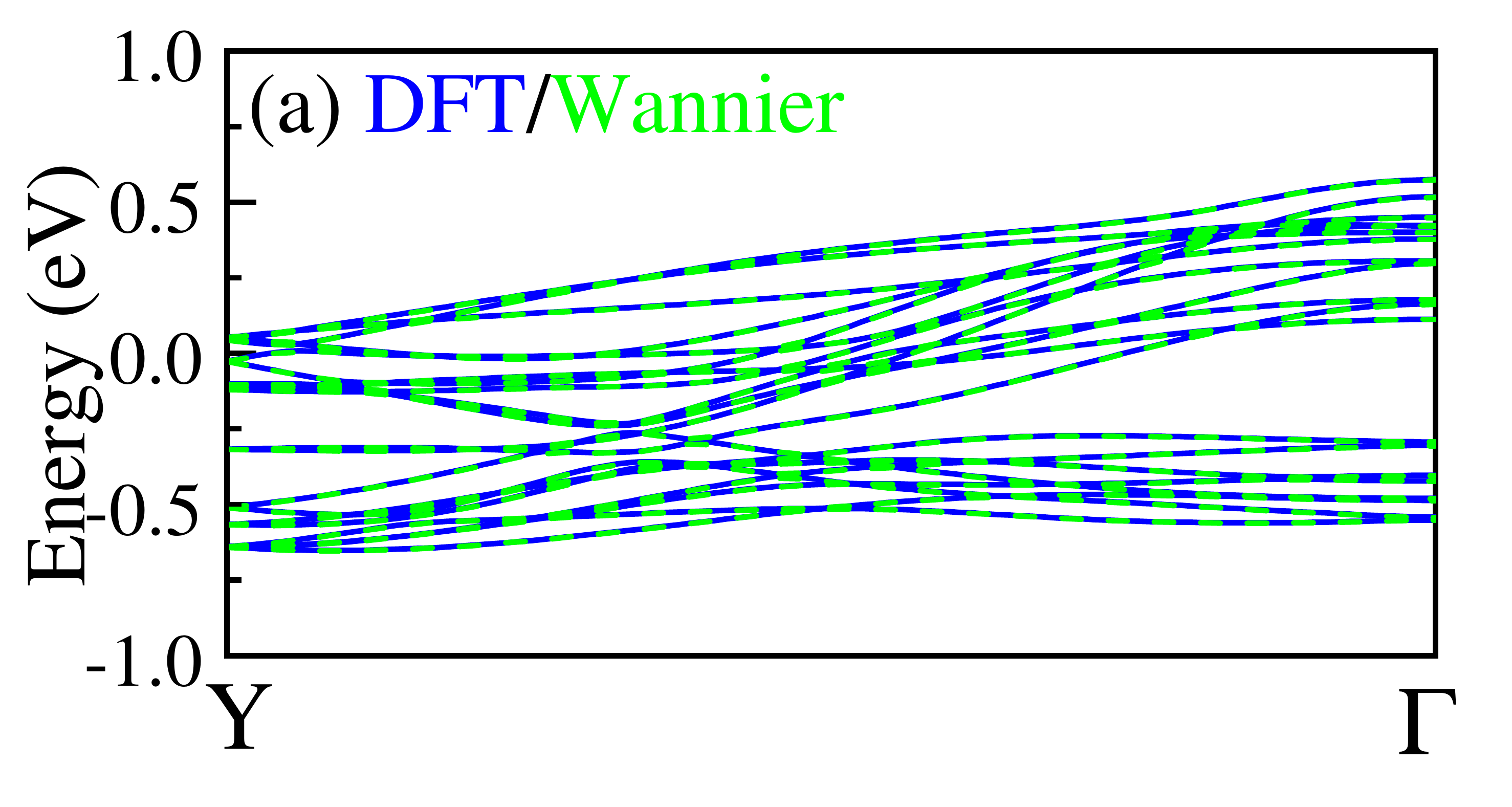}
\includegraphics[width=0.48\textwidth]{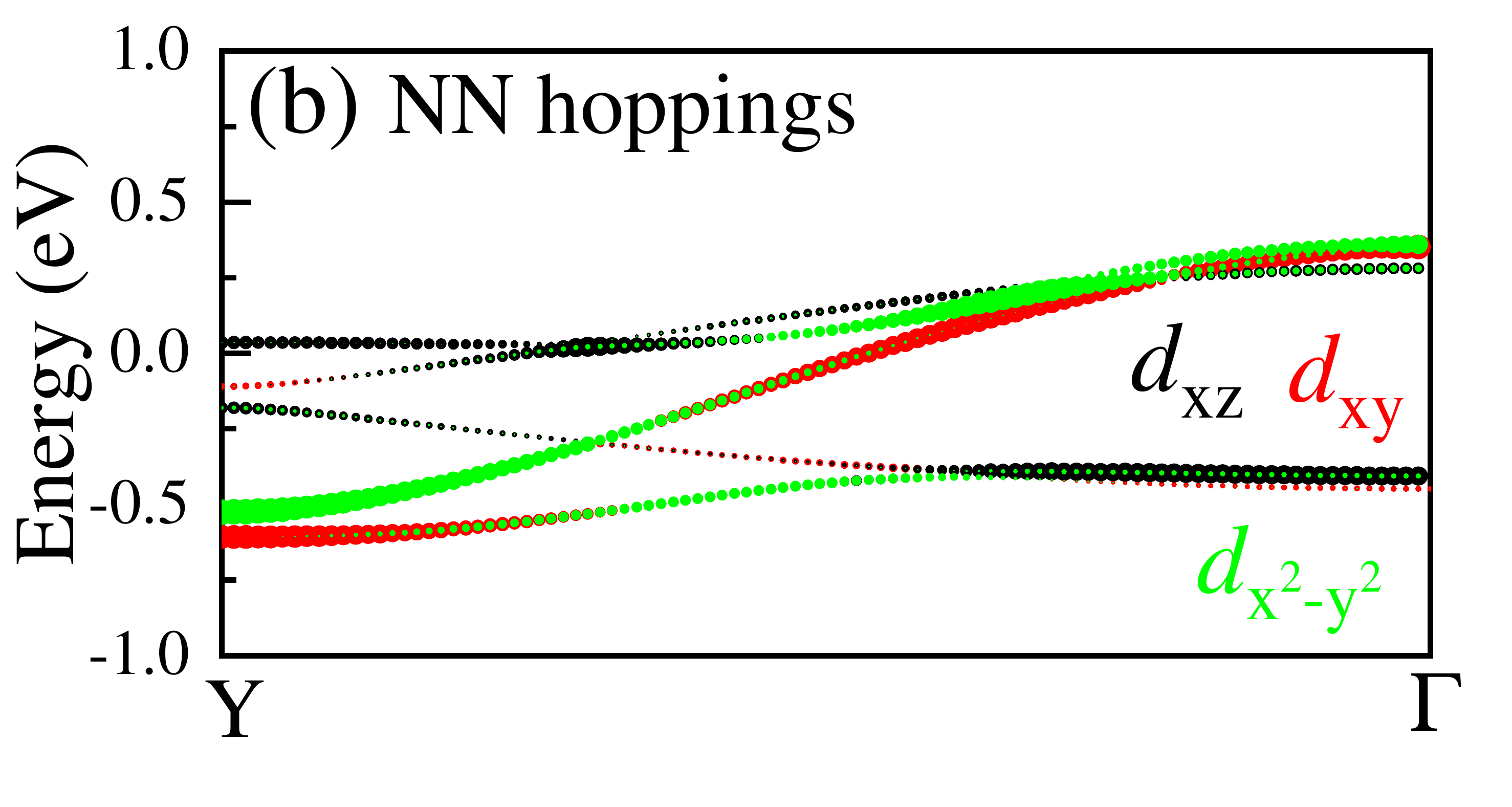}
\caption{(a) DFT and Wannier bands of Ba$_2$FeS$_3$ (HP) along the FeS$_4$ chain direction ($b$-axis). (b) Three-orbital tight-binding model with nearest-neighbor hoppings along the $b$ axis. The BZ points are Y = (0, 0.5, 0) and $\Gamma$ = (0, 0, 0). Note that Y is in scaled units, corresponding to the units of $2\pi$/b.}
\label{Fig14}
\end{figure}

Here, we focus only on the iron chain since the intrachain coupling is the key aspect to understand the physical properties of Ba$_2$FeS$_3$ (HP). Thus, we used the MLWFs to fit the DFT bands along the $b$-axis (Y-$\Gamma$), corresponding to the quasi-one-dimensional electronic characteristics of Ba$_2$FeS$_3$ (HP), as displayed in Fig.~\ref{Fig14}(a). Based on the Wannier fitting results, we deduced the hopping parameters and on-site matrix.

Considering the computational limitation of the DMRG method, we constructed a three-orbital model involving the orbital basis $d_{xz}$, $d_{x^2-y^2}$ and $d_{xy}$ for the iron chain, readjusted to properly fit the band structure after reducing the original five orbitals to three. The three-orbital tight-binding bands agree qualitatively well with the DFT band structure, as displayed in Fig.~\ref{Fig14}(b).

Based on the Wannier fitting, we obtained four on-site matrices for the four Fe atoms in a unit cell, using the basis \{$d_{z^2}$, $d_{xz}$, $d_{yz}$, $d_{x^2-y^2}$, $d_{xy}$\}.

\begin{equation}
\begin{split}
t_{onsite}^1 =
\begin{bmatrix}
          d_{z^2}   &    d_{xz}         &      d_{yz}      &      d_{x^2-y^2}   &     d_{xy}   \\
          3.812  	&     -0.063 	    &      0.000	   &       0.075	    &      0.000	\\
         -0.063     &      3.628	    &      0.000	   &       0.183	    &      0.000	\\
          0.000	    &      0.000	    &      3.509	   &       0.000	    &     -0.054	\\
          0.075	    &      0.183	    &      0.000	   &       3.644	    &      0.000	\\
          0.000	    &      0.000	    &     -0.054	   &       0.000	    &      3.618	
\end{bmatrix},\\
\end{split}
\label{onsite1}
\end{equation}

\begin{equation}
\begin{split}
t_{onsite}^2 =
\begin{bmatrix}
          3.812	    &     -0.063	    &      0.000	   &       0.075	   &       0.000	  \\
         -0.063	    &      3.628	    &      0.000	   &       0.183	   &       0.000	  \\
          0.000	    &      0.000	    &      3.509	   &       0.000	   &      -0.054	  \\
          0.075     &      0.183	    &      0.000	   &       3.644	   &       0.000	  \\
          0.000	    &      0.000	    &     -0.054	   &       0.000	   &       3.618	
\end{bmatrix},\\
\end{split}
\label{onsite2}
\end{equation}

\begin{equation}
\begin{split}
t_{onsite}^3 =
\begin{bmatrix}
          3.812	    &      0.063	    &      0.000	   &       0.075	   &       0.000	  \\
          0.063	    &      3.628	    &      0.000	   &      -0.183	   &       0.000	  \\
          0.000	    &      0.000	    &      3.509	   &       0.000	   &       0.054	  \\
          0.075     &     -0.183	    &      0.000	   &       3.644	   &       0.000	  \\
          0.000	    &      0.000	    &      0.054	   &       0.000	   &       3.618		
\end{bmatrix},\\
\end{split}
\label{onsite3}
\end{equation}

\begin{equation}
\begin{split}
t_{onsite}^4 =
\begin{bmatrix}
          3.812	    &      0.063	    &      0.000	   &       0.075	   &       0.000	  \\
          0.063	    &      3.628	    &      0.000	   &      -0.183	   &       0.000	  \\
          0.000	    &      0.000	    &      3.509	   &       0.000	   &       0.054	  \\
          0.075     &     -0.183	    &      0.000	   &       3.644	   &       0.000	  \\
          0.000	    &      0.000	    &      0.054	   &       0.000	   &       3.618	
\end{bmatrix}.\\
\end{split}
\label{onsite4}
\end{equation}

Furthermore, we also obtained four nearest-neighbors hopping matrices along the $b$-axis, corresponding to the four Fe atoms in a unit cell.

\begin{equation}
\begin{split}
t_{\vec{b}}^1 =
\begin{bmatrix}
          0.057  	&     -0.094 	    &      0.071	   &       0.048	    &      0.011	\\
         -0.094     &     -0.003	    &     -0.019	   &       0.020	    &     -0.083	\\
         -0.071     &      0.019	    &     -0.016	   &       0.055	    &     -0.132	\\
          0.048	    &      0.020	    &     -0.055	   &       0.169	    &      0.022	\\
         -0.011	    &      0.083	    &     -0.132	   &      -0.022	    &      0.172	
\end{bmatrix},\\
\end{split}
\label{hopping1}
\end{equation}

\begin{equation}
\begin{split}
t_{\vec{b}}^2 =
\begin{bmatrix}
          0.057  	&     -0.094 	    &     -0.071	   &       0.048	    &     -0.011	\\
         -0.094     &     -0.003	    &      0.019	   &       0.020	    &      0.083	\\
          0.071     &     -0.019	    &     -0.016	   &      -0.055	    &     -0.132	\\
          0.048	    &      0.020	    &      0.055	   &       0.169	    &     -0.022	\\
          0.011	    &     -0.083	    &     -0.132	   &       0.022	    &      0.172	
\end{bmatrix},\\
\end{split}
\label{hopping2}
\end{equation}

\begin{equation}
\begin{split}
t_{\vec{b}}^3 =
\begin{bmatrix}
          0.057  	&      0.094 	    &      0.071	   &       0.048	    &     -0.011	\\
          0.094     &     -0.003	    &      0.019	   &      -0.020	    &     -0.083	\\
         -0.071     &     -0.019	    &     -0.016	   &       0.055	    &      0.132	\\
          0.048	    &     -0.020	    &     -0.055	   &       0.169	    &     -0.022	\\
          0.011	    &      0.083	    &      0.132	   &       0.022	    &      0.172	
\end{bmatrix},\\
\end{split}
\label{hopping3}
\end{equation}

\begin{equation}
\begin{split}
t_{\vec{b}}^4 =
\begin{bmatrix}
          0.057  	&      0.094 	    &     -0.071	   &       0.048	    &      0.011	\\
          0.094     &     -0.003	    &     -0.019	   &      -0.020	    &      0.083	\\
          0.071     &      0.019	    &     -0.016	   &      -0.055	    &      0.132	\\
          0.048	    &     -0.020	    &      0.055	   &       0.169	    &      0.022	\\
         -0.011	    &     -0.083	    &      0.132	   &      -0.022	    &      0.172	
\end{bmatrix}.\\
\end{split}
\label{hopping4}
\end{equation}

As shown above, there are some non-zero off-diagonal elements in the on-site matrices, indicating the constructed MLWFs orbitals are not exactly orthogonal
to one other. Hence, we introduced a unitary matrix transformation
to reconstruct the effective on-site and hopping matrices:

\begin{equation}
\begin{split}
U =
\begin{bmatrix}
          0.881  	&     -0.246        &      0.000	   &      -0.406	    &      0.000	\\
          0.131     &     -0.696	    &      0.000	   &       0.706	    &      0.000	\\
          0.000     &      0.000	    &      0.925	   &       0.000	    &     -0.381	\\
          0.456     &      0.675	    &      0.000	   &       0.580 	    &      0.000	\\
          0.000	    &      0.000	    &      0.381	   &       0.000	    &      0.925	
\end{bmatrix},\\
\end{split}
\label{unit}
\end{equation}

As discussed in the main text, the Ba$_2$FeS$_3$ (HP) is a quasi-one-dimensional system, where the physical properties are primarily contributed by the intrachain coupling. Hence, we just considered one iron chain and NN hopping in our DMRG calculations. The reconstructed on-site and hopping matrices are:

\begin{equation}
\begin{split}
t_{onsite}^1 =
\begin{bmatrix}
          d_{z^2}   &     d_{xz}        &      d_{yz}      &      d_{x^2-y^2}   &     d_{xy}   \\
          3.841  	&      0.000 	    &      0.000	   &       0.000	    &      0.000	\\
          0.000     &      3.428	    &      0.000	   &       0.000	    &      0.000	\\
          0.000	    &      0.000	    &      3.487	   &       0.000	    &      0.000	\\
          0.000	    &      0.000	    &      0.000	   &       3.814	    &      0.000	\\
          0.000	    &      0.000	    &      0.000	   &       0.000	    &      3.640	
\end{bmatrix},\\
\end{split}
\label{onsiteR}
\end{equation}

\begin{equation}
\begin{split}
t_{\vec{b}}^1 =
\begin{bmatrix}
          0.098  	&      0.119 	    &      0.035	   &      -0.006	    &     -0.005	\\
          0.119     &      0.012	    &     -0.011	   &       0.045	    &      0.080	\\
         -0.035     &      0.011	    &     -0.082	   &       0.087	    &     -0.028	\\
         -0.006	    &      0.045	    &     -0.087	   &       0.112	    &     -0.018	\\
          0.005	    &     -0.080	    &     -0.028	   &       0.018	    &      0.238	
\end{bmatrix}.\\
\end{split}
\label{hoppingR}
\end{equation}

Here, we used the three orbitals \{$d_{xz}$, $d_{x^2-y^2}$, $d_{xy}$\} in our calculations, corresponding to the electronic density per orbital $n = 4/3$. As explained before, this electronic density is widely used in the context of iron low-dimensional compounds with DMRG technology, where the ``real''  iron is in a valence Fe$^{\rm 2+}$, corresponding to six electrons in five orbitals per site~\cite{osmp1,Luo:prb10}. In our DMRG calculations, the on-site and hopping matrices are:
\begin{equation}
\begin{split}
t_{onsite} =
\begin{bmatrix}
           d_{xz}      &      d_{x^2-y^2}   &     d_{xy}   \\
           3.428	   &       0.000	    &      0.000	\\
           0.000	   &       3.814	    &      0.000	\\
           0.000	   &       0.000	    &      3.640	
\end{bmatrix},\\
\end{split}
\label{onsiteF}
\end{equation}

\begin{equation}
\begin{split}
t_{\gamma\gamma'} =
\begin{bmatrix}
          0.012     &   0.045  &   0.080	   	       \\
          0.045     &   0.112  &  -0.018	   	       \\
         -0.080	    &   0.018  &   0.238	
\end{bmatrix}.\\
\end{split}
\end{equation}

\subsection{B. DMRG results for $L = 24$}

\begin{figure}
\centering
\includegraphics[width=0.48\textwidth]{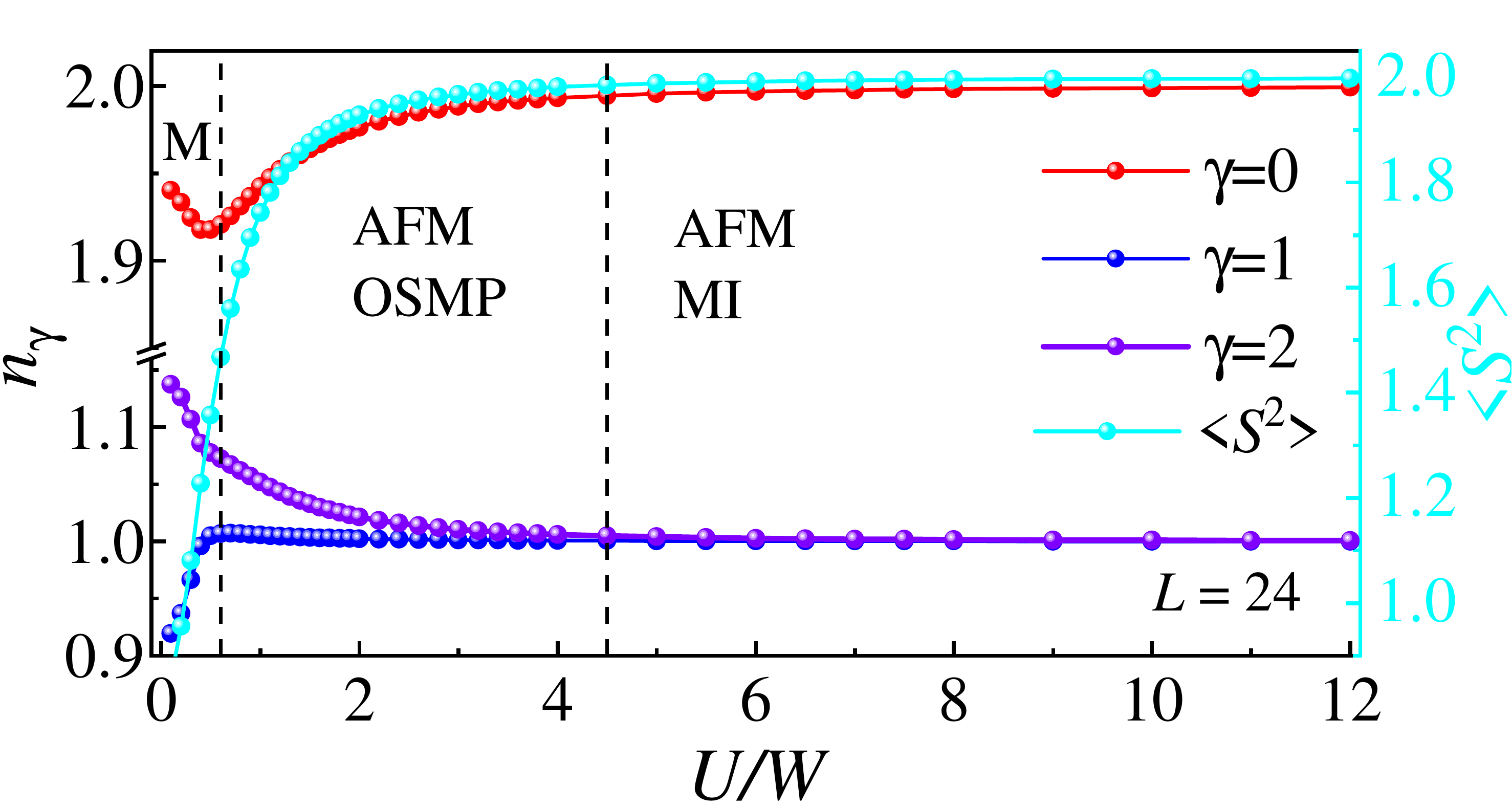}
\caption{Orbital-resolved occupation number $n_{\gamma}$, mean-value of the total spin-squared $\langle{{S}}^2\rangle$, at different values of $U/W$ and $J_{\rm H}/U = 1/4$. Here, we used a $24$-sites cluster chain with nearest-neighbor hoppings for four electrons in three orbitals.}
\label{Fig15}
\end{figure}

As displayed in Fig.~\ref{Fig15}, we show the site-averaged occupancy of different orbitals $n_{\gamma}$ vs $U/W$ for $L = 24$, at the typical value of $J_H/U$. Those results are similar to the results of $L = 16$ (Fig.~\ref{Fig7}), indicating that our results are robust against changes in $L$ (small size effects).

\begin{references}
\bibitem{Jcontext} Here, using a $1\times2\times1$ cell, we calculated $J_1$ by comparing the difference of energies of FM and AFM order along the $b-$axis while the interchain couplings were regarded as FM. The $J_2$ and $J_3$ were calculated by various magnetic configurations in a unit cell.
\end{references}
\end{document}